\documentclass[final,3p,times]{elsarticle}
\usepackage{graphicx}
\usepackage{caption}
\usepackage{subcaption}
\usepackage{amsmath,amssymb}
\journal{Journal of Computational Physics}

\begin{document}
\begin{frontmatter}
\title{Fokker Planck kinetic modeling of suprathermal $\alpha$ particles in a fusion plasma}


\author[label1]{B. E. Peigney\corref{cor1}}
\cortext[cor1]{benjamin.peigney@mines-paris.org}
\author[label1]{O. Larroche}
\author[label2]{V. Tikhonchuk}
\address[label1]{CEA/DIF, BP 12, 91680 Bruy\`eres le Ch\^atel, France}
\address[label2]{University Bordeaux -- CNRS -- CEA, CELIA 33405 Talence Cedex, France}

\begin{abstract}
We present an ion kinetic model describing the ignition and burn of the deuterium-tritium fuel of inertial fusion targets. The analysis of the underlying physical model enables us to develop efficient numerical methods to simulate the creation,  transport and  collisional relaxation of fusion reaction products ($\alpha$-particles) at a kinetic level.  A two-energy-scale approach leads to a self-consistent modeling of the coupling between suprathermal $\alpha$-particles and the thermal bulk of the imploding plasma. This method provides an accurate numerical treatment of energy deposition and transport processes involving suprathermal particles. The numerical tools presented here are validated against known analytical results. This enables us to investigate the potential role of ion kinetic effects on the physics of ignition and thermonuclear burn in inertial confinement fusion schemes. 
\end{abstract}

\end{frontmatter}

\section{Purpose of the study}\label{sec1}

Inertial confinement fusion (ICF) is a process of energy production obtained from the nuclear fusion reaction between deuterium (D) and tritium (T) ions. It is a promising and abundant energy source for future power plants. The fusion reactions $D+T \rightarrow \alpha + n + 17.56$\,MeV take place in a hot and dense plasma compressed and heated by intense laser radiation. The thermonuclear burn of the  deuterium-tritium (DT) fuel is supported by energetic $\alpha$-particles, which are created by fusion reactions at the energy 3.52\,MeV. Those suprathermal particles subsequently transfer their energy to the fresh fuel through Coulomb collisions. 


In the case of Inertial Confinement Fusion \cite{LIN981, Atzeni}, a spherical DT shell is compressed to densities of the order of a few hundred g/cc by the ablation pressure. Fusion reactions start in a central zone characterized by a density $\rho \sim 50$\,g.cm$^{-3}$ and a high ''ignition`` temperature $T\approx 7-10$\,keV. The surrounding shell is 10 times colder than the hot spot ($T\approx 0.7$\,keV). The density of the central ''hot spot`` is such that  the mean free path $\lambda_\alpha$ of fast $\alpha$-particles is roughly equal to the hot spot radius $R$\cite{FRA744}. This allows the self-heating of the hot spot fuel which serves as a spark that subsequently burns the surrounding colder and denser shell.

The design of ICF targets and the interpretation of ICF experiments rely on numerical simulations based on hydrodynamic Lagrangian codes where  kinetic effects are only considered as corrections included in the transport coefficients \cite{LIN981, Atzeni}. The fluid description is relevant if the mean free path of plasma particles, namely electrons and ions, is smaller than the characteristic length scale. Although this condition is reasonably fulfilled during the implosion stage, it does not apply to fast particles, in particular to fusion products near the ignition threshold. Thus, an accurate kinetic modeling is required.

The purpose of the present work is to propose an ion-kinetic description of suprathermal fusion products, treated self-consistently  with the ion-kinetic modeling of the thermal imploding plasma. 
The difficulty lies in the coupling of ion populations characterized by two different energy scales: 
\begin{itemize}
\item{Thermal particles $D$,$T$, which form the bulk of the imploding plasma and whose kinetic energy is in the keV range.}
\item{Suprathermal $\alpha$-particles, created at 3.52 MeV by fusion reactions.}
\end{itemize}
Such a strong disparity in energy scales makes it difficult to build viable kinetic models of fusion reactions.

Existing ion kinetic codes can describe the implosion of DT targets in sub-ignition conditions  \cite{CAS91A, VID95A, LAR03A}, but the energy release from the fusion reactions is not accounted for in a self-consistent manner. Several simplified methods compatible with hydrodynamic codes have been developed. Haldy and Ligou \cite{LIGOU1} apply the moment method to model ion energy deposition  in a hot and dense homogeneous plasma, but only a stationary case has been considered. A variety of methods based on diffusion models applied to charged-particle transport problems have also been developed. Those methods are of considerable interest, since results on energy deposition profiles can be obtained with a low computational effort. Nevertheless, diffusion methods rely on the assumption that the fast particle mean free path is smaller than the characteristic scale length of the energy deposition zone. This hypothesis does not hold for a typical ICF target near ignition. Corman et al \cite{COR754} derive a multi-group diffusion model from the Fokker-Planck equation to describe fast ion transport in a fusion plasma. However, they introduce heuristically a flux limiter in order  to prevent unphysical behavior when particle flux approaches the free-streaming limit. Pomraning  \cite{POM33} develops a more sophisticated flux limiter scheme based on the Chapman-Enskog expansion. However, the  flux limited diffusion smoothes artificially energy deposition profiles, especially in situations where ion sources are localized \cite{HON931}. This may lead to significant errors in the calculation of ignition thresholds and energy gains. Such diffusion models are employed in all major present-day fluid codes because of their compatibility with the underlying hydrodynamic module.

Several exact methods can be employed to solve the Fokker-Planck equation in a general way, but they are too much time consuming. Monte Carlo  algorithms are applied to model charged particle transport in Refs. \cite{LAP1, SENT1}. In such an approach, distribution functions are represented by a sum of Dirac measures. Monte Carlo particles are characterized by their numerical weight, their position and their velocity. Those quantities evolve in time according to the Vlasov-Fokker-Planck equation while the tracking of Monte Carlo particles is performed through the spatial mesh. The accuracy of Monte Carlo methods is proportional to $N^{-1/2}$, $N$ being the number of Monte Carlo particles, so that $N\gg 1$ and variance reduction techniques are usually employed to reduce numerical noise. A significant deficiency of Monte Carlo methods for the investigation of kinetic effects is that the tails of the distribution functions are not described  accurately. Moreover, the coupling between suprathermal particles and the thermal bulk is usually treated in a rough manner, by removing the  suprathermal particles that are slowed down below a given energy threshold and injecting the removed particles in the thermal bulk. Therefore, the thermalization process is not described with a sufficient precision.

$S_n$ methods are also used to solve the Fokker-Planck equation deterministically. They are based on the determination of the angular flux of suprathermal particles at a set of discrete directions, each one associated with a quadrature weight \cite{MEHL1, KIL1, Duclous}. Although they are more accurate than diffusion methods and can be extended to highly anisotropic particle distribution functions, the weakly collisional limit is not described accurately and the  thermalization process is treated approximately with the same strategy as in Monte Carlo methods. $S_n$ methods are usually used to simulate neutron transport and require high computational efforts.  For the application of $S_n$ methods to suprathermal $\alpha$-particles transport, we refer to Ref. \cite{HON931}.

In the present paper we develop a kinetic modeling of suprathermal fusion products in the thermal imploding plasma.  We extend the existing code \textsc{FPion} \cite{CAS91A, VID95A, LAR03A} so as to treat $\alpha$-particles, for which \textit{two scales of energy} are considered, namely a suprathermal and a thermal one. Since the developments made to reach this goal have been substantial, they have actually lead to the creation of an entirely new kinetic code called  \textsc{Fuse}  for \emph{\textsc{FPion} Upgrade with two Scales of Energy}. This code is able to investigate  kinetic effects related to fusion reaction products on the ignition of the hot spot and on the subsequent propagation of the thermonuclear burn wave through the dense fuel. We present here the numerical methods specially designed for the kinetic modeling of $\alpha$-particles and their validation in several representative tests. Simulations are preformed for a typical ICF DT target, assuming a spherical symmetry in configuration space and axial symmetry in velocity space around the mean velocity. Distribution functions thus depend on one space variable  (radius) and two velocity components (radial and azimuthal or perpendicular), depending on the chosen parametrization.  

The paper is organized as follows: firstly, we present in Sec.\,\ref{sec2} the Vlasov-Fokker-Planck modeling of the fast $\alpha$-particle transport and collisional relaxation. A specific formalism, based on a two-scale approach with respect to energy is then introduced in Sec.\,\ref{sec3}. It provides a self-consistent modeling of the coupling between suprathermal and thermal plasma  species.  Section\,\ref{sec4} presents the algorithms devised to solve the two-scale coupling. A finite volume method is applied to the Fokker-Planck equation governing the suprathermal $\alpha$-particle distribution function. Fast algorithms are then specially designed to solve the discretized model efficiently. Section\,\ref{sec5} presents some numerical results regarding the $\alpha$-particle distribution function evolution and its coupling with the thermal bulk. We show how the methods developed here provide a refined description of the thermalization process. Simulations are carried out in conditions relevant for typical ICF targets. Conclusions are finally presented in Sec.\,\ref{sec6}.

\section{Physical model for the transport and collisional relaxation of $\alpha$-particles}\label{sec2}

Once created by fusion reactions, suprathermal $\alpha$-particles are transported through an inhomogeneous plasma and slowed down through  Coulomb collisions with electrons and thermal ions D and T. Besides, pressure gradients give rise to an electrostatic field  $\vec{\mathcal{E}}(\vec r,t)$ that may accelerate or decelerate $\alpha$-particles. To give an accurate description of the particle transport, as well as the non-local energy and momentum exchange that occur between $\alpha$-particles and the thermal bulk, a kinetic modeling based on the Vlasov-Fokker-Planck equation is required.
 
\subsection{Vlasov-Fokker-Planck equation for the $\alpha$-particles}\label{sec21}

The distribution function $f_\alpha(\vec r,\vec v,t)$ of $\alpha$-particles characterized by a  charge $Z_\alpha e$ and a mass $m_\alpha$ is governed by the Vlasov-Fokker-Planck equation:
\begin{equation}
\label{eq:vfp_alpha} 
\displaystyle\frac{\partial f_\alpha}{\partial t}+\vec v \cdot \frac{\partial f_\alpha}{\partial \vec r} +
\frac{Z_\alpha e \vec{\mathcal{E}}}{m_\alpha}\cdot\frac{\partial f_\alpha}{\partial \vec v} =\sum_{i}\left.\frac{\partial
f_\alpha}{\partial t}\right|_{\alpha i} + \left.\frac{\partial f_\alpha}{\partial t}\right|_{\alpha e} + \left.\frac{\partial f_\alpha}{\partial t}\right|_{\rm fuse}.
\end{equation}
The first two terms at the right hand side of this equation describe the collisional relaxation of $\alpha$-particles:
\begin{itemize}
\item{$\partial f_\alpha/\partial t|_{\alpha e}$ stands for the collisions of $\alpha$-particles with electrons,}
\item{$\sum_{i}\partial f_\alpha/\partial t|_{\alpha i}$ describes the collisions of $\alpha$-particles with thermal ion species. Since thermal species densities are significantly higher than the fast $\alpha$-particle density (at least at the beginning of the ignition and burn processes), non-linear term corresponding to fast-$\alpha$/fast-$\alpha$ scattering is neglected. The coupling between the thermalized $\alpha$ particles and the suprathermal ones is naturally included.}


\end{itemize}

We focus now on the collisional part of Eq.\,\eqref{eq:vfp_alpha}. The Vlasov part of the equation modeling the transport in space and the acceleration due to the electrostatic field is considered separately in Sec.\,\ref{sec4}.
In a fully ionized plasma such as the one considered here, large angle scattering are much less likely than the net large-angle deflection due to a cumulative effect of many small-angle collisions that the projectile experiences along its path \cite{ROS573}. Each of the collision terms in right hand side of Eq.\,\eqref{eq:vfp_alpha} can then be expressed  as a Fokker-Planck operator in velocity space, which amounts essentially to an advection-diffusion form. More precisely, the slowing down of $\alpha$-particles on a thermal ion species $i$ can be written as:


\begin{equation}
\label{eq:fp_alpha_i}
\left.\frac{\partial f_\alpha}{\partial t}\right|_{\alpha i} = 4\pi\Gamma_{\alpha i}
\frac{\partial}{\partial \vec v}\cdot\left(\frac{m_\alpha}{m_i}f_\alpha \frac{\partial
\mathcal{S}_i}{\partial \vec v}-\nabla^2_v \mathcal{T}_i\cdot\frac{\partial f_\alpha}{\partial \vec v}\right),
\end{equation}
where $\mathcal{S}_i$ and $\mathcal{T}_i$ are the so-called Rosenbluth potentials \cite{ROS573} associated to the target ions $i$. They are defined by a set of Poisson equations in velocity space:
\begin{equation}
\label{eq:poisson_rosenbluth}
\Delta_v \mathcal{S}_i = f_i, \qquad \Delta_v \mathcal{T}_i = \mathcal{S}_i.
\end{equation}
The coefficient $\Gamma_{\alpha i} = (4\pi Z_\alpha^2 Z_i^2 e^4/m_\alpha^2)\ln\Lambda_{\alpha i}$ is proportional to the Coulomb logarithm $\ln\Lambda_{ij}$ (for any species $i,j$ including electrons) related to the Coulomb potential screening  and taking quantum effects into account: $\Lambda_{ij}=\lambda_D/\max\{\lambda_{\rm bar},\rho_\bot\}$. The Debye length
$$ \lambda_D =\left(4\pi n_e e^2/T_e+\sum_{j=1}^n 4\pi n_j Z_j^2 e^2/T_j\right)^{-1/2} $$
depends on the temperature $T_j$, which is expressed in energy units. $T_j$ is related to the thermal ion distribution function $f_j$  by the relation:
$$ T_j = \frac{m_j}{3n_j}  \int (v-V_j)^2 f_j(\vec v)\, d^3v,$$
where $n_j = \int f_j(\vec v)\, d^3 v$ is the density of ion species $j$ and $\vec V_j = n_j^{-1} \int \vec v f_j(\vec v)\, d^3v$ is their mean velocity. The characteristic lengths $\rho_\bot$ and $\lambda_{\rm bar}$ are the  classical and quantum impact parameters:
$$ \rho_\bot =  Z_a Z_be^2/m_{ij}u_{ij}^2, \qquad \lambda_{\rm bar} =  \hbar/m_{ij}u_{ij}$$
where $m_{ij}=m_i m_j/(m_i+m_j)$ is the reduced mass and $u_{ij}= \sqrt{3}(T_i/m_i + T_j/m_j)^{1/2}$ is an average relative velocity between the particle species $i$ and $j$. The Coulomb logarithm is thus a particular function of hydrodynamic quantities. It is symmetric with the respect of particle species, $\Lambda_{ij}=\Lambda_{ji}$, which is related to the energy and momentum conservation during the collision.

The effect of electrons on the slowing down of $\alpha$-particles is modeled by another Fokker-Planck term, in which the electron distribution function is approximated by a Maxwellian characterized by a density $n_e$, a mean velocity $\vec u_e$ and a temperature $T_e$:


\begin{equation}
\label{eq:fp_alpha_e}
\left.\frac{\partial f_\alpha}{\partial t}\right|_{\alpha e} = \displaystyle\frac{1}{\tau_{e\alpha}} \frac{\partial}{\partial \vec v}\cdot \left[(\vec v - \vec u_e) f_\alpha(\vec v) + \frac{T_e}{m_\alpha} \frac{\partial f_\alpha}{\partial v_\alpha}(\vec v) \right],
\end{equation}
where $\tau_{e\alpha}$ is a characteristic $e-\alpha$ collision time defined by:
\begin{equation}
\label{eq:tauei}
\tau_{e\alpha} = \displaystyle\frac{3}{4\sqrt{2\pi}}\displaystyle\frac{m_\alpha T_e^{3/2}}{n_e  Z_\alpha^2  e^4 m_e^{1/2} \ln\Lambda_{\alpha e} }.
\end{equation}
Equation \eqref{eq:fp_alpha_e} is obtained by a truncated expansion of the full ion-electron Fokker-Planck operator with respect to the small constant $\epsilon = (m_e/m_i)^{1/2} \sim 0.022$ \cite{CAS91A, LAR03A}. 



The last term in \eqref{eq:vfp_alpha} stands for the creation of $\alpha$-particles by fusion reactions. The source term is supposed to be isotropic and is given by:
\begin{equation}
\label{eq:evol_fdh_source}
\left.\frac{\partial f_\alpha}{\partial t}\right|_{\rm fuse}= \mathcal{R}_{DT}(\vec r,t)\frac{\delta(v-v_h)}{4\pi v^2},
\end{equation}
where $v_h= 1.3\times 10^9$\,cm.s$^{-1}$ is the initial velocity of suprathermal $\alpha$-particles whose initial energy is 3.52\,MeV. $\mathcal{R}_{DT}$ is the fusion reaction rate expressed as a function of the distribution functions of D and T, respectively:
\begin{equation}
\label{eq:tau_reac}
\mathcal{R}_{DT}(\vec r,t) = n_D n_T \langle\sigma v\rangle_{DT} = \int\int f_D(\vec r,\vec v_D,t)\, f_T(\vec r,\vec v_T,t)\,|\vec v_D - \vec v_T|\,\sigma_{DT}(|\vec v_D - \vec v_T|)\,d^3v_D d^3v_T.
\end{equation}
The distribution functions $f_D$ and $f_T$ are solutions of the Vlasov-Fokker-Planck equation  written on the deuterium and tritium species, respectively, and they are not necessarily Maxwellian functions. Integrals in Eq.\,\eqref{eq:tau_reac} are taken over the three-dimensional velocity space.

\subsection{Dealing with electrons}\label{sec22}

Since the characteristic time of the considered problem is close to the ion-ion collision time $\tau_{ii}>>1/\omega_{pe}$, $\omega_{pe}$ being the electron plasma frequency, and the characteristic length is of the order of the ion collisional mean free path $\lambda_i>>\lambda_{De}$, $\lambda_{De}$ being the electron Debye length, the quasi-neutrality assumption is relevant. We then have:

\begin{equation}
\label{eq:quasineut}
n_e =\sum_i Z_i n_i+Z_\alpha n_\alpha^{ST}, \qquad\vec V_e = \sum_i Z_i n_i \vec V_i+Z_\alpha n_\alpha^{ST} V_\alpha^{ST},
\end{equation}


where the contribution of suprathermal $\alpha$-particles is naturally included, $n_\alpha^{ST},V_\alpha^{ST}$ being the density and mean velocity of fast $\alpha$-particles respectively.

Besides, due to a very small ratio of the masses of electrons and ions, the electron equilibration time $\tau_{ee}$ is significantly smaller than the mean ion-ion collision time $\tau_{ii}$. According, for example to \cite{BRA65A},  we have the following ordering of characteristic times: $\tau_{ee} \sim \epsilon \tau_{ii}$.  As a consequence, the electron kinetic equation reduces to a fluid equation. Only an equation for the temperature (or, equivalently, the energy density) is actually needed since  the electron density and velocity are known from the quasi-neutrality conditions (\ref{eq:quasineut}).


In the one-dimensional spherical problem considered here, the electron energy density $W_e$ is governed by the following conservation equation~:
\begin{equation}
\frac{\partial W_e}{\partial t} +\frac{1}{r^2}\frac{\partial}{\partial r}
\left(r^2u_eW_e\right) + \frac{1}{r^2}\frac{\partial}{\partial r}(r^2u_e)P_e   - \frac{1}{r^2}\frac{\partial}{\partial r}\left(r^2\kappa_e
\frac{\partial T_e}{\partial r}\right)= \sum_{j=1}^n \frac{3n_j}{2\tau_{ej}}(T_j-T_e) +
\left.\frac{\partial W_e}{\partial t}\right|_{\rm rad}
\label{eqTe}
\end{equation}
where $\kappa_e$ is Spitzer's thermal conductivity \cite{SPI532} in the presence of several ion species (see also \cite{LAR93} and Appendix in \cite{CHE979}), the collision time $\tau_{ej}$ has been defined in Eq.\,\eqref{eq:tauei} where  $\alpha$ is replaced by the considered ion species $j$. The electron energy density $W_e$ and pressure $P_e$ are given by an equation of state  taking into account Fermi degeneracy \cite{LAR03A}. 


The last term on the right hand side of \eqref{eqTe} accounts for the radiation losses of electrons.

\subsection{Relative importance of electrons and ions on the slowing down of $\alpha$-particles}\label{sec23}


3.52 MeV $\alpha$-particles are created in fusion reactions isotropically, in the system of reference associated with the thermal bulk. Then, they are slowed down through Coulomb collisions with electrons, according to Eq. \,\eqref{eq:fp_alpha_e}, and with thermal ions, according to Eq.\,\eqref{eq:fp_alpha_i}. The relative importance of electrons and ions on the slowing down of $\alpha$-particles can be estimated by retaining only the dynamical friction terms from the Fokker-Planck equations \eqref{eq:fp_alpha_e} and \eqref{eq:fp_alpha_i}. The ratio $R_{i/e}$ between the ion slowing down and the electron one can thus be approximated by:
$$ R_{i/e} = \left.\frac{\partial f_\alpha}{\partial t}\right|_{\alpha i}\left/\right.\left.\frac{\partial f_\alpha}{\partial t}\right|_{\alpha e} \sim \frac{T_e^{3/2}}{v^3 m_e^{1/2}m_i} \sim \frac{T_e^{3/2}}{v^3 m_i^{3/2} \epsilon}. $$
The ratio $R_{i/e}$ is thus defined by a characteristic threshold velocity: 
\begin{equation}
\label{eq:vcoup}
v_{c}= \epsilon^{-1/3}(T_e/m_i)^{1/2},
\end{equation}

so that $R_{i/e} \sim (v_{c}/v)^3$.

The beginning of the slowing-down of $\alpha$ particles is thus governed nearly exclusively by electrons. Then, as $v \sim v_{c}$, the effect of ions and electrons on the $\alpha$ relaxation become comparable. Eventually, the final stage of $\alpha$-particle thermalization is essentially influenced by collisions with thermal ions.  Supposing $T_i \sim T_e$, we have the following estimate $v_{c} \sim \epsilon^{-1/3} v_i^{th} \sim 3.6\, v_i^{th}$, $v_i^{th}$ being the typical thermal velocity of D and T ions. The effect of thermal ions on the $\alpha$ relaxation dominates when the $\alpha$ velocity is below $v_{c}\sim 3.6 \,v_i^{th}$. We shall refer to such $\alpha$-particles as ''moderately suprathermal``. 

\section{Two-component description of the $\alpha$ distribution function}\label{sec3}
\subsection{Physical discussion}\label{sec31}

From the previous discussion, we know that 3.52 MeV $\alpha$-particles are firstly slowed down essentially by electrons. The first stage of the $\alpha$ slowing down is thus described by:
\begin{equation}
\label{eq:fsure}
\left.\frac{\partial f_\alpha}{\partial t}\right|_{\rm coll} = \frac{1}{\tau_{\alpha e}} \frac{\partial}{\partial \vec v}\cdot \left[(\vec v - \vec u_e) f_\alpha(\vec v) + \frac{T_e}{m_\alpha} \frac{\partial f_\alpha}{\partial \vec v}(\vec v) \right].
\end{equation}
When $v >> u_e$, the dynamic friction term (first term on the right hand side of (\ref{eq:fsure})) dominates so that the $\alpha$ distribution evolves with respect to:
\begin{equation}
\label{eq:fsurebis}
\left(\frac{\partial f_\alpha}{\partial t}\right)_{coll}  \approx \frac{1}{\tau_{\alpha e}}  \frac{1}{v^2}\frac{\partial}{\partial v}\cdot \left[v^3  f_\alpha(v) \right].
\end{equation}
The stationary solution of (\ref{eq:fsurebis}) behaves as $f_\alpha \sim 1/v^3$, where $v$ is suprathermal $\alpha$-particle velocity. Consequently, as long as fast $\alpha$-particles remain far from the thermal velocity region, their distribution function varies smoothly over the whole suprathermal velocity region. The associated velocity scale $v_\alpha^{ST}$, defined by:
\begin{equation}
\label{eq:v_ast}
v_\alpha^{ST} \sim f_\alpha^{ST}/{\displaystyle\frac{\partial f_\alpha^{ST}}{\partial v}},
\end{equation}
is in particular greater than the target thermal velocity $v_i^{th}$.

Then, when slowed down $\alpha$-particles get closer to the thermal region but still remain suprathermal, thermal ions  tend to dominate the end of the relaxation process, which is then governed by the equation:
\begin{equation}
\label{eq:modele_1d}
\left.\frac{\partial f_\alpha}{\partial t}\right|_{\rm coll} =\sum_{i} 4\pi\Gamma_{\alpha i}\frac{\partial}{\partial\vec v}\cdot\left(\frac{m_\alpha}{m_i}f_\alpha \frac{\partial\mathcal{S}_i}{\partial \vec v}\right),
\end{equation}
where only the dynamical friction term is retained for the present discussion. We shall deal with the diffusion part separately. Qualitatively, one can consider that the distribution function of the thermal target species $i$ appears highly localized in velocity space, from the suprathermal $\alpha$-particle point of view. One thus can write:
$ f_i(\vec v) = n_i \delta^3(\vec v)$ (assuming that the mean velocity is zero). Besides, the divergence with respect to velocity that appears on the right hand side of Eq.\,\eqref{eq:modele_1d} can be expanded as follows:
$$
\frac{\partial}{\partial \vec v}\cdot \left(\frac{\partial\mathcal{S}_i}{\partial \vec v}f_\alpha\right)\simeq \frac{\partial\mathcal{S}_i}{\partial \vec v}\cdot \frac{\partial f_\alpha}{\partial \vec v} + f_\alpha \Delta_v \mathcal{S}_i.
$$
Using the approximation $f_i(\vec v) = n_i \delta^3(\vec v)$, which is valid from the suprathermal $\alpha$-particle point of view, the first Rosenbluth potential associated to the target ions $i$ can be calculated explicitly: $\mathcal{S}_i(v) \sim -n_i/(4\pi v)$. Then, by calculating its derivative, the slowing down of $\alpha$ particles can be modeled by:
\begin{equation}
\label{eq:modele_1dbis}
\left.\frac{\partial f_\alpha}{\partial t}\right|_{\rm coll} =\sum_{i} 4\pi\Gamma_{\alpha i} \frac{m_\alpha}{m_i}\left( \frac{\partial f_\alpha}{\partial \vec v} \cdot \frac{n_i}{4\pi v^2}\vec{e}_v + f_\alpha f_i  \right).
\end{equation}
The two terms on the right hand side of Eq.(\ref{eq:modele_1dbis}) have a clear physical sense. The first term $\sim \partial f_\alpha/\partial \vec v$ varies slowly and smoothly far from the thermal velocity region. It can be characterized by a suprathermal velocity scale $v_{\alpha}^{ST}$, which is greater than the typical thermal ion velocity $v_i^{th}=(T_i/m_i)^{1/2}$. Actually, the term $\sim \displaystyle\frac{n_i}{4\pi v^2} \displaystyle\frac{\partial f_\alpha}{\partial \vec v}$ corresponds to a conservative convection towards $v=0$. The associated convective rate $\displaystyle\frac{n_i}{4\pi v^2}$ increases as $v$ tends to $0$ so that the solution of:
\begin{equation}
\label{eq:convec_mod}
\left(\frac{\partial f_\alpha}{\partial t}\right)_{coll} =\sum_{i} 4\pi\Gamma_{\alpha i} \frac{m_\alpha}{m_i}\left[ \frac{\partial f_\alpha}{\partial \vec v} \cdot \frac{n_i}{4\pi v^2}\vec{e}_v \right]
\end{equation}
tends to a constant $f_0$ corresponding to the stationary state of (\ref{eq:convec_mod}). The part of the $\alpha$ distribution driven by (\ref{eq:convec_mod}) is then stretched and smoothed out as it approaches the thermal velocity region.


 The second term $\sim f_\alpha f_i$ appears highly localized in the thermal region of velocity space and behaves qualitatively as a $\delta$-function from the suprathermal $\alpha$-particle point of view. This term  actually leads to the formation of a condensate of width $v_i^{th} \ll v_{\alpha}^{ST}$.

This qualitative analysis shows intuitively how the \textit{two-component feature} of the $\alpha$ distribution function builds up. It is made of a superposition of two components evolving on two different velocity scales, namely:
\begin{itemize}
\item{a suprathermal component, fed by fusion reactions and evolving on a large velocity scale, greater than the target thermal velocity.}
\item{A thermal component, corresponding to the thermalized part of the $\alpha$ distribution function, evolving on the same velocity scale as the thermal bulk of the plasma. Note that this component is not fully thermalized since the source term is proportional to $\sum_i 4\pi \Gamma_{\alpha i} f_i$. There remains a final stage of collisional relaxation between the thermal components of D,T and $\alpha$ ions respectively.} 
\end{itemize}

\begin{figure}[!h]
\centering
{\includegraphics[width=0.4\textwidth]{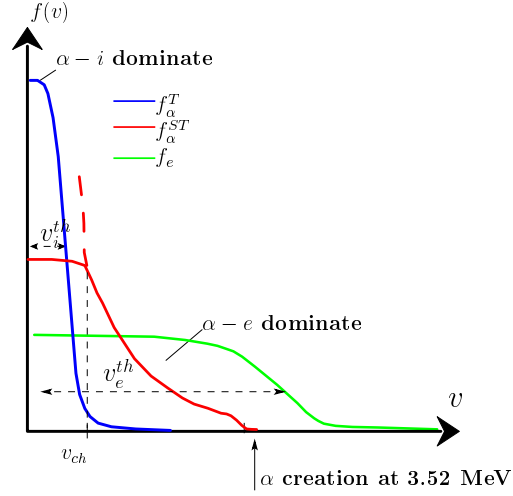}}
\caption{\label{fig:model1d} Schematic representation of the collisional relaxation of suprathermal $\alpha$-particles on thermal target ions $i$.
The suprathermal component of the $\alpha$ distribution (red) varies on a velocity scale $v_\alpha^{ST} \gg v_i^{th}$. The  electron distribution function (green) has a Maxwellian shape with a characteristic width $v_e^{th} \gg v_i^{th}$. The thermal ion component (blue) varies on the thermal ion energy scale $\sim v_i^{th}$. The contrast between the thermal and suprathermal scales has been reduced artificially for the sake of clarity.
}
\end{figure}

Figure \ref{fig:model1d} illustrates schematically those processes. From this phenomenological discussion, we can draw a more formal and more rigorous description of the slowing-down which naturally leads to the building of a new multi-scale algorithm solving the initial problem given by Eq.\,\eqref{eq:vfp_alpha}.

\subsection{Splitting of the Fokker-Planck operator}\label{sec32}

From the previous analysis, it seems natural to write the $\alpha$ distribution function as follows:
\begin{equation}
\label{eq:split_fd}
f_\alpha(\vec v,t) =  f_\alpha^{ST}(\vec v,t)+f_\alpha^{T}(\vec v,t),
\end{equation}
where: $f_\alpha^{ST}$ designates the suprathermal component. It is defined on a large velocity domain, spreading to the MeV range. Its typical velocity variation scale $v_{\alpha}^{ST}$ is greater than the thermal ion velocity $v_i^{th}$; $f_\alpha^{T}$ is the thermal component. It is localized in the region of velocity space corresponding to target thermal ion distribution functions and vanishes in the suprathermal velocity domain. The component $f_\alpha^{T}$ is designed to describe accurately the final stage of thermalization of the slowed down $\alpha$-particles. This final relaxation occurrs on a velocity scale $\sim v_i^{th}$.

Let us emphasize that two components defined in Eq.\,\eqref{eq:split_fd} do exist in the whole velocity space, the relevant physical quantity being the full $\alpha$ distribution function $f_\alpha(\vec v,t)$. 


The idea is then to deal with each component separately. The original Fokker-Planck operator given in Eq.\,\eqref{eq:fp_alpha_i} is then transformed into a \textit{system of two coupled equations} governing the two components $f_\alpha^{ST}$ and $f_\alpha^{T}$, respectively:
\begin{eqnarray}
\label{eq:system_ST_T}
&& \left.\partial_t f_\alpha^{ST}\right|_{\alpha i} =  \Gamma_{\alpha i} \frac{n_i}{v^2} \partial_{v} f_\alpha^{ST} - n_i\Gamma_{\alpha i} f_\alpha^{ST} \frac{\delta(v)}{v^2}, \nonumber\\
&& \left.\partial_t f_\alpha^T\right|_{\alpha i} = 4\pi\Gamma_{\alpha i} \partial_{\vec v}\cdot\left(f_\alpha^{T} \partial_{\vec v}\mathcal{S}_i\right) + 4\pi\Gamma_{\alpha i}  f_i  f_\alpha^{ST}(v=0).
\end{eqnarray}

The above equations are written in the system of reference associated with the thermal ions.

System \eqref{eq:system_ST_T} describes the coupling between the suprathermal component and the thermal one, the coupling function being $\sim f_\alpha^{ST} f_i$, which is subtracted from the equation on the suprathermal component $f_\alpha^{ST}$ and appears as a source term in the equation governing the thermal component $f_\alpha^T$. The coupling function can actually be approximated for each of the components of the $\alpha$ distribution function in two different ways, depending on the considered velocity scale:


\begin{itemize}
\item{From the suprathermal component point of view, we have $f_\alpha^{ST} f_i \sim n_i f_\alpha^{ST} \delta^3(\vec v)$  since thermal target ions appear highly localized. The first Rosenbluth potential $\mathcal{S}_i$ associated to thermal ions can then be approximated by its temperature-vanishing form.}
\item{From the point of view of the thermal component, we can consider $f_\alpha^{ST} f_i \sim f_\alpha^{ST}(0) f_i$ since the suprathermal component is almost constant on the thermal velocity scale $v_i^{th}$. The term $\sim f_\alpha^{ST}(0) f_i$ appears as a source term for the thermal component. It corresponds to a feeding by the suprathermal component.}
\end{itemize}
In Eq.\,\eqref{eq:system_ST_T}, we have disregarded the process corresponding to a feeding of the suprathermal component by the thermal one, which could be the case if we modeled large angle collisions, such as  $\alpha^{ST} + D \to \alpha + D^{ST}$. Such collisions would build up a suprathermal component for species $D$ and $T$. This could be naturally included in the formalism that we describe here, but this is a process of second order since the probability of large angle scattering is $\sim 1/\ln\Lambda$ times smaller than the pitch-angle collisions modeled by the Fokker-Planck operator.

\subsection{Diffusion part of the Fokker-Planck operator}\label{sec33}

We study now the effect of the second term on the right hand side of  Eq.\,\eqref{eq:fp_alpha_i} corresponding to a diffusion in velocity:
\begin{equation}
\label{eq:fp_alpha_i2}
\left.\frac{\partial f_\alpha}{\partial t}\right|_{\alpha i} = -\sum_{i} 4\pi\Gamma_{\alpha i}\frac{\partial}{\partial \vec v}\cdot\left( \nabla^2_v \mathcal{T}_i\cdot\frac{\partial f_\alpha}{\partial \vec v}\right).
\end{equation}
$\mathcal{T}_i$ is the second Rosenbluth potential associated to the thermal target ions. The notation $\nabla^2_v (\,.\,)$ stands for the Hessian $\partial^2_{\alpha\beta}(\,.\,)$.
Let us define the field $\vec J_{\alpha i}$, representing the slowing-down current of $\alpha$-suprathermal particles:
\begin{equation}
\label{eq:jperp_def}
\vec J_{\alpha i}= -\sum_{i} 4\pi\Gamma_{\alpha i} \nabla^2_v \mathcal{T}_i \partial f_\alpha/\partial \vec v,
\end{equation}
Using the Dirac-function approximation for the thermal target distribution functions, we can approximate  $\mathcal{T}_i$ by its temperature-vanishing form, $\mathcal{T}_i(v) \sim -n_i v/(8\pi)$. The approximation is relevant from the suprathermal component point of view. The Hessian $\nabla^2_v \mathcal{T}_i$ can then be calculated explicitly:
\begin{equation}
\label{eq:hess_1}
\nabla^2_v \mathcal{T}_i \sim -\frac{n_i}{8\pi v} \left(\mbox{Id}-\frac{\vec v \otimes \vec v}{v^2}\right).
\end{equation}
By taking advantage of a polar representation of the velocity $\vec v=v\vec{e}_v$, where $(\vec e_v, \vec e_\theta)$ is the polar local basis of velocity space, the Hessian \eqref{eq:hess_1} simplifies to:
\begin{equation}
\label{eq:hess_2}
\nabla^2_v\mathcal{T}_i \sim -\frac{n_i}{8\pi v}\,\vec{e}_\theta\otimes \vec{e}_\theta\,.
\end{equation}
The slowing down current defined in  Eq.\,\eqref{eq:jperp_def} expresses the diffusion in velocity associated to the slowing-down process. It is essentially transverse, that is, perpendicular to the local velocity $\vec v$. Therefore, one can write:
\begin{equation}
\label{eq:jperp_approx}
\vec J_{\alpha i} \sim -\displaystyle\frac{\Gamma_{\alpha i}}{2} \frac{n_i}{v^2} \frac{\partial f_\alpha}{\partial\theta} \vec{e}_\theta.
\end{equation}
The diffusive slowing-down current is thus highly anisotropic in velocity space and it intensifies as $\alpha$-particles approaches the thermal bulk region of velocity space. Qualitatively, the collisional relaxation of $\alpha$-particles on  thermal target ions is thus characterized by:
\begin{itemize}
\item{a pure advection in velocity space at a constant rate, modeled by Eq.\,\eqref{eq:modele_1d}, which tends to accumulate $\alpha$-particles in the thermal ion velocity region.}
\item{An anisotropic diffusion in velocity space, expressed by Eq.\,\eqref{eq:jperp_approx}, which tends to make the distribution isotropic when slowed-down $\alpha$-particles get closer to the final stage of thermalization.}
\end{itemize}

\section{Algorithms for the transport and collisional relaxation of fast fusion products}\label{sec4}

In this section, we present the numerical methods developed to solve  Eq.\,\eqref{eq:vfp_alpha} and Eq.\,\eqref{eq:system_ST_T}. Those equations govern the time evolution of suprathermal $\alpha$-particles. Firstly, we show how to deal with  the two-component nature of the $\alpha$ distribution function. We then develop a finite volume approach to discretize the equation on the $\alpha$ suprathermal component. An efficient explicit algorithm is then applied to model the time evolution of the suprathermal component with relatively low computational time. We finally present how to simulate accurately the complete thermalization process of $\alpha$-particles.

\subsection{Co-existence of two velocity grids}\label{sec41}

The two-component nature of the $\alpha$ distribution function naturally leads to the co-existence of two velocity grids, namely:
\begin{itemize}
\item{A suprathermal grid, designed to represent the evolution of the suprathermal component of the $\alpha$ distribution function $f_\alpha^{ST}$. It covers a large domain in velocity, extending to the range $v\simeq v_h\simeq 1.3\,10^{9}$ cm/s, which is the velocity corresponding to the  $\alpha$ particles created by fusion reactions. Moreover, since the suprathermal component varies smoothly, we can use a relatively coarse grid to discretize it. $f_\alpha^{ST}$ varies significantly on a velocity scale $v_\alpha^{ST} \gg v_i^{th}$, so that the suprathermal grid resolution is typically of the order of one thermal  velocity $v_i^{th}$.}
\item{A thermal grid, on which the thermal component of the $\alpha$ distribution $f_\alpha^{T}$ is discretized. This grid is designed to capture the final stage of collisional relaxation of the almost-thermalized component of the $\alpha$ distribution on the other thermal ion species D and T. This process entails a velocity resolution much smaller than the local thermal velocity scale $v_i^{th}$. The thermal grid makes use of a cylindrical parametrization $(v_r,v_\bot)$ inherited from the code \textsc{Fpion}\cite{LAR93}.}
\end{itemize}

\vspace{0cm}
\begin{figure}[!h]
\centering
{\includegraphics[width=0.5\textwidth]{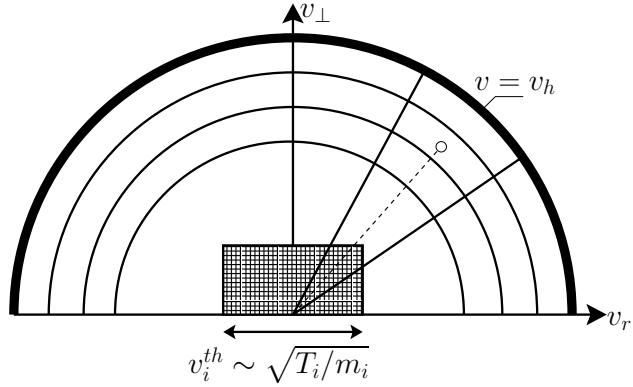}}
\caption{\label{fig:2mesh} Schematic representation of the two velocity grids used to model the $\alpha$ suprathermal and thermal components respectively. The suprathermal component evolves on the coarse polar grid, covering a wide domain extending to the MeV region. The thick shell of width $\sim T_i$ corresponds to the source term due to fusion reactions. The thermal component evolves on the small and refined cylindrical grid. Both meshes are centered on the mean local bulk velocity $V_0\sim V_e \sim V_i$. Velocity space is characterized by an axial symmetry around the axis $\vec v_r$.
}
\end{figure}
The two grids that are shown in figure \ref{fig:2mesh} are centered on the local mean bulk velocity $V_0(r)$, which is close to the mean electron velocity $V_e(r)$. By using two grids specially-tailored to capture the variations of each component, it is possible to build an efficient algorithm modeling the two components of the $\alpha$ distribution. 

\subsection{Dimensionless form of the Vlasov-Fokker-Planck equation}\label{sec42}

For numerical purposes, we write the Vlasov-Fokker Planck equation governing the evolution of the suprathermal component of the $\alpha$ distribution function $f_\alpha^{ST}$ in a dimensionless form, based on a specified unit system given in Table \ref{tabunit}. It is  chosen to manipulate numbers that are close to unity. This prevents computational errors caused by under or overflow floating numbers.   As it was shown in Eq.\,\eqref{eq:jperp_approx}, the collision term between suprathermal $\alpha$-particles and ions takes a simple form expressed in polar coordinates. The slowing down currents are co-linear with the local polar basis vectors $\vec{e}_v,\vec{e}_\theta$ of velocity space. In the spherical one-dimensional geometry considered here, it thus seems natural to parametrize the suprathermal distribution function as $f^{ST}_\alpha(r,v,\theta,t)$, with two velocity components $\vec v = v \cos\theta \,\vec{e}_r + v\sin\theta \, \vec{e}_\bot$.  
Then, the dimensionless equation governing $f^{ST}_\alpha$ reads:
\begin{eqnarray}
&& \frac{\partial f^{ST}_\alpha}{\partial t} +  v\,\cos\theta\,\frac{\partial f^{ST}_\alpha}{\partial r} + \frac{ 
\mathcal{E}_\alpha}{A_\alpha} \cos\theta \frac{\partial  f^{ST}_\alpha}{\partial v} = \sum_{i}  \widetilde{\Gamma}_{\alpha i} \displaystyle\frac{\partial}{\partial \vec v} \cdot \, \left[\frac{n_i}{ v^2} \left(\frac{A_\alpha}{A_i} f_\alpha^{ST} \vec{e}_v + \frac{1}{2}\frac{\partial f_\alpha^{ST}}{\partial\theta}\vec{e}_\theta \right)\right] \nonumber \\
&&  + \frac{1}{\widetilde{\tau}_{e\alpha}}\displaystyle\frac{\partial}{\partial \vec v} \cdot \,\left[(\vec v - \vec {u_e})f_\alpha^{ST} + \frac{T_e}{A_\alpha}\frac{\partial}{\partial \vec v}f_\alpha^{ST}\right] - \sum_{i=D,T,\alpha} 4\pi\widetilde{\Gamma}_{\alpha i}\frac{A_\alpha}{A_i} f^{ST}_\alpha f_i^T + \mathcal{R}_{DT}(\vec r,t)\frac{\delta(v-v_h)}{4\pi v^2},
\label{eq:FP_supra_discr}
\end{eqnarray}
where the normalized constant $\widetilde{\Gamma}_{\alpha i}= (4\pi Z_\alpha^2 Z_\beta^2/A_i^2)\ln\Lambda_{\alpha i}$ and the
effective electrostatic field $\mathcal{E}_i $ applied to ions of species $i$ is defined by the following expression:



\begin{equation}\label{Ei}
\mathcal{E}_i = - (Z_i/\widetilde{n}_e)\,\partial \widetilde{P}_e/\partial r.
\end{equation}
Here, $\widetilde{n}_e$ and $\widetilde{P}_e$ are the dimensionless electron density and pressure, respectively, and
$$ \widetilde{\tau}_{e\alpha} = \frac{3\sqrt{\pi}A_\alpha T_e^{3/2}}{2\epsilon \sqrt{2}Z_\alpha^2n_e\ln\Lambda_{\alpha e}}$$ 
is the dimensionless electron-ion collision time.

\begin{table}[!h]
\caption{Units defined from reference values of the particle density $n_0$ and
particle thermal energy $T_0$.}
\label{tabunit}
\begin{center}
\begin{tabular}{l l}
\hline\noalign{\smallskip}
Quantity & Unit\\
\noalign{\smallskip}\hline\noalign{\smallskip}
density & $n_0$ (arbitrary reference value) \\
thermal energy & $T_0$ (arbitrary reference value) \\
time & $\tau_0 = T_0^{3/2}m_p^{1/2}/4\pi e^4n_0$ \\
length & $\lambda_0 = (T_0/m_p)^{1/2}\tau_0 = T_0^2/4\pi e^4n_0$ \\
velocity & $v_0 = (T_0/m_p)^{1/2} = \lambda_0/\tau_0$ \\
distribution function & $f_0 = n_0/v_0^3$ \\
first Rosenbluth pot. & $\mathcal{S}_0 = n_0/v_0$ \\
second Rosenbluth pot. & $\mathcal{T}_0 = n_0v_0$ \\
electric field ($\mathcal{E}_i$) & $\mathcal{E}_0=m_pv_0^2/\lambda_0=m_p\lambda_0/\tau_0^2$ \\
heat flux & $Q_0 = n_0T_0^{3/2}/m_p^{1/2}$ \\
\noalign{\smallskip}\hline
\end{tabular}
\end{center}
\end{table}

Let us consider the third term on the right hand side of \eqref{eq:FP_supra_discr}. From the point of view of suprathermal $\alpha$-particles, it can be approximated by:
\begin{equation}
\label{eq:third_term}
\sum_{i} 4\pi\widetilde{\Gamma}_{\alpha i}\frac{A_\alpha}{A_i} f^{ST}_\alpha f_i \simeq 4\pi\sum_{i} \widetilde{\Gamma}_{\alpha i}\frac{A_\alpha}{A_i} f^{ST}_\alpha n_i \delta^3(\vec v),
\end{equation}
supposing that  $v \gg v_i^{th},V_0$. The term \eqref{eq:third_term} is thus highly peaked with respect to velocity in the thermal component region and leads to the formation of a thermalized condensate that cannot be described on the coarse suprathermal grid.  That justifies our approach of subtracting this singular term from \eqref{eq:FP_supra_discr}, so that the variations of $f_\alpha^{ST}$ remain everywhere smooth and may be described on the suprathermal grid. The term  \eqref{eq:third_term} is then re-introduced as a \textit{feeding term} in the equation governing the thermal component, so that the original Fokker-Planck equation governing the complete $\alpha$ distribution function $f_\alpha=f_\alpha^{ST}+f_\alpha^{T}$ is recovered. 

To solve the full Vlasov-Fokker-Planck equation \eqref{eq:FP_supra_discr}, we use the same general splitting scheme as in the code \textsc{FPion}, namely we treat the advection, the acceleration  and the collisional stages separately. We describe now the method developed to solve the collisional part of \eqref{eq:FP_supra_discr}.

\subsection{Discretization of the collisional term}\label{sec43}

The collisional part of (\ref{eq:FP_supra_discr}) can be written as:
\begin{equation}
\label{eq:collis_st_discr1}
\left.\frac{\partial f_{\alpha}^{st}}{\partial t}\right|_{\mbox{coll}} = \frac{1}{v^2}\frac{\partial}{\partial v}\left(v^2 J^v\right) + \frac{1}{v\sin\theta}\frac{\partial}{\partial \theta}\left(\sin\theta \,J^\theta\right),
\end{equation}
where the polar components of the slowing down current $\vec J$ are given by:
\begin{equation}
\label{eq:Jv_st_discr}
J^v = f^{ST}_\alpha \left(\frac{v}{\tau_{e \alpha}}+\widetilde{\Gamma}_{\alpha i}\frac{A_\alpha}{A_i} \displaystyle\frac{n_i}{v^2}\right) +\frac{1}{\widetilde{\tau}_{e\alpha}} \frac{T_e}{A_\alpha}\frac{\partial f_\alpha^{ST}}{\partial v},
\end{equation}
and 
\begin{equation}
\label{eq:Jth_st_discr}
J^\theta = \displaystyle\frac{1}{v}\displaystyle\frac{\partial f_\alpha^{ST}}{\partial\theta}\left(\widetilde{\Gamma}_{\alpha i}\displaystyle\frac{n_i}{2v}+\displaystyle\frac{1}{\widetilde{\tau}_{e\alpha}} \displaystyle\frac{T_e}{A_\alpha} \right),
\end{equation}
The slowing-down current $\vec J$ takes the general advection-diffusion form in velocity space:
\begin{equation}
\begin{pmatrix}J^v \\ J^\theta\end{pmatrix} = f\begin{pmatrix}u_v \\ u_\theta\end{pmatrix}
+ \begin{pmatrix}K^{vv} & K^{v\theta} \\ K^{\theta v} & K^{\theta\theta}\end{pmatrix}\cdot
\begin{pmatrix}\displaystyle\frac{\partial f}{\partial v} \\ 
\\
\displaystyle\frac{1}{v}\displaystyle\frac{\partial f}{\partial\theta}\end{pmatrix}
\end{equation}
where the components of the tensors $u$ and $K$ are related to the Rosenbluth potentials $\mathcal{S}$ and $\mathcal{T}$ (associated to the target ion species) as follows:
\[
\begin{pmatrix}u^v \\ 
\\
u^\theta\end{pmatrix} = \begin{pmatrix}\displaystyle\frac{\partial \mathcal{S}}{\partial v} \\ 
\\
\displaystyle\displaystyle\frac{1}{v}\displaystyle\frac{\partial \mathcal{S}}{\partial\theta}\end{pmatrix}\quad\mbox{and}\quad
\begin{pmatrix}K^{vv} & K^{v\theta} \\ 
\\
K^{\theta v} & K^{\theta\theta}\end{pmatrix} =
\begin{pmatrix}
\displaystyle\frac{\partial^2\mathcal{T}}{\partial v^2} & \displaystyle\frac{\partial}{\partial v}\left(\displaystyle\frac{1}{v}\displaystyle\frac{\partial \mathcal{T}}{\partial\theta}\right) \\
\\
\displaystyle\frac{\partial}{\partial v}\left(\displaystyle\frac{1}{v}\frac{\partial \mathcal{T}}{\partial\theta}\right) & \displaystyle\frac{1}{v^2}\displaystyle\frac{\partial^2\mathcal{T}}{\partial\theta^2} + \displaystyle\frac{1}{v}\displaystyle\frac{\partial \mathcal{T}}{\partial v}
\end{pmatrix}
\]
which reduces to:
\begin{equation}
\label{eq:uK_def}
\begin{pmatrix}u^v \\ u^\theta\end{pmatrix} = \begin{pmatrix} v/\widetilde{\tau}_{e\alpha} + \sum_{i=D,T} \widetilde{\Gamma}_{\alpha i} n_i/ v^2\\ 0 \end{pmatrix}\quad\mbox{and}\quad
\begin{pmatrix}K^{vv} & K^{v\theta} \\ K^{\theta v} & K^{\theta\theta}\end{pmatrix} =
\begin{pmatrix}  T_e/\widetilde{\tau}_{e\alpha} A_\alpha & 0\\  0 & \sum_{i=D,T} \widetilde{\Gamma}_{\alpha i}  n_i/(2v)
\end{pmatrix}\,.
\end{equation}
Note the simplifications implied by using a polar parametrization of velocity space: the dynamical friction coefficient $\vec u$ is indeed co-linear with the radial velocity basis vector $\vec e_v$ and the diffusion tensor is diagonal in the basis $\vec e_v,\vec e_\theta$.

\begin{figure}[!h]
\centering
{\includegraphics[scale=0.4]{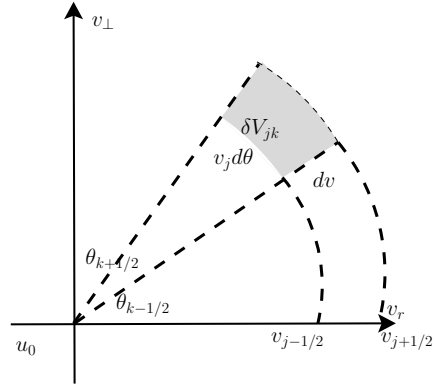}}
\caption{\label{polarmesh} The suprathermal velocity grid.
}
\end{figure}

We then integrate \eqref{eq:collis_st_discr1} with respect to velocity on a given cell $\delta V_{kj}$ of the polar suprathermal velocity grid, subscripts $k$ and $j$ referring to the $\theta$ and $v$ directions respectively (see figure \ref{polarmesh}).

The cell $\delta V_{kj}$ is defined by its boundaries $\theta_{k-\frac{1}{2}}$, $\theta_{k+\frac{1}{2}}$ and $v_{j-\frac{1}{2}}$, $v_{j+\frac{1}{2}}$, for  $1\le k\le k_{max}$ and  $1\le j\le j_{max}$.  We call $f_{kj}^n = f_\alpha^{ST}(v=v_j,\theta=\theta_k,t=t_n)$ the value of the suprathermal distribution function in the cell $\delta V_{kj}$ at time $t_n$.  Integrating Eq.\,\eqref{eq:collis_st_discr1} over the cell area $\delta V_{kj}$,  we obtain the following conservative discretized form:
\begin{equation}
\label{eq:scheme_st_1}
\frac{f_{kj}^{n+1}-f_{kj}^n}{\Delta t}  = \frac{1}{v_j^2}\frac{v_{j+1/2}^2J^v_{kj+1/2}-v_{j-1/2}^2J^v_{kj-1/2}}{2\delta v^3_j}
+ \frac{3v_j\delta v_j}{2\delta v^3_j}\frac{\sin\theta_{k+1/2}J^\theta_{k+1/2j}-\sin\theta_{k-1/2}J^\theta_{k-1/2j}}{\delta\mu_k}
\end{equation}
where discrete elementary volumes are defined by:
$$ \delta v^3_j = v_{j+\frac12}^3-v_{j-\frac12}^3, \quad \delta v_j = v_{j+\frac12}-v_{j-\frac12},\quad 
\delta \mu_k = \cos\theta_{k+\frac12}-\cos\theta_{k-\frac12}. $$
The centered radial velocity $v_j$ that appears in Eq.\,\eqref{eq:scheme_st_1} is defined as $ v_j = (v_{j+\frac12}+v_{j-\frac12})/2$.
In those notations, the discrete volume of the cell $\delta V_{kj}$  is given by :
$$
\delta V_{kj} = \int_{\delta V_{kj}}2\pi v^2\sin\theta \,dv\,d\theta = \frac{4\pi}{3}\delta v^3_j \delta \mu_k.
$$
Besides, a straightforward centered-difference and explicit discretization of the slowing-down current leads to:
\begin{eqnarray}
&& J^v_{kj+1/2} = \frac{u^v_{kj+1/2}}{2}(f^n_{kj+1}+f^n_{kj}) -\frac{K^{vv}_{kj+1/2}}{\delta v_{j+1/2}}(f^n_{kj+1}-f^n_{kj}) \label{eq:Jv_kj_discr1} \\
&& J^\theta_{k+1/2j} = \frac{K^{\theta\theta}_{k+1/2j}}{v_j\delta\theta_{k+1/2}}(f^n_{k+1j}-f^n_{kj}), \label{eq:Jth_kj_discr1}
\end{eqnarray}
where the slowing-down coefficient $u$ and the diffusion coefficients $K$ are explicitly given by \eqref{eq:uK_def} as  functions of velocity. The time varying coefficients in \eqref{eq:uK_def} involving thermal ions and electrons are evaluated at the previous time step $t=t_n$.

\subsection{A Locally Split Explicit scheme}\label{sec44}
\subsubsection{Need for an explicit approach}\label{sec441}

The slowing-down and diffusion coefficients given in Eq.\,\eqref{eq:uK_def} are thus very inhomogeneous in velocity space, being highly peaked in magnitude near the thermal component region. Besides, the diffusion term is strongly anisotropic (essentially transverse) outside of the thermal component region. In such a situation, the usual implicit schemes may involve the solution of a very large and ill-conditioned linear system that will only give an approximated solution  of the non-stationary problem. In this section, we demonstrate how it is possible to take advantage of the strong inhomogeneity  of the slowing down current to build an efficient and  simple explicit scheme that describes the non-stationary $\alpha$ distribution function time evolution naturally. This approach stems from ideas that were introduced in \cite{LAR07}.

The Von Neumann stability condition for the scheme \eqref{eq:scheme_st_1} in the case of constant homogeneous slowing-down coefficient $u$ and  diffusion tensor $K$ reads as:
\begin{equation}
\label{eq:stabcond1}
(u \,\delta t)^2 \le 2\mbox{Tr}(K)\,\delta t \le \delta v^2,
\end{equation}
where  $\delta v$ is the velocity mesh size. When the slowing-down coefficient $u$ and diffusion tensor $K$ are inhomogeneous (which is the case for our problem), we can apply \eqref{eq:stabcond1} \textit{locally} in each cell $\delta V_{jk}$ of the suprathermal polar velocity grid. Besides, since the scheme \eqref{eq:scheme_st_1} is bi-dimensional and parametrized in polar coordinates, \eqref{eq:stabcond1} actually leads to two stability conditions, corresponding to the radial direction $v$ and the angular direction $\theta$, respectively.

Treating these directions separately, the stability condition for \eqref{eq:scheme_st_1} can be written for a given cell $\delta V_{jk}$ as:
\begin{itemize}
\item{in the radial $v$ direction:}
\begin{equation}
\label{eq:stabcond_v}
\left(\displaystyle\frac{u^v_j\delta t}{\delta v_j}\right)^2 \le \frac{2(K^{vv}_{j})\delta t}{\delta v_j^2} \le 1
\end{equation}
\item{in the angular $\theta$ direction:} 
\begin{equation}
\label{eq:stabcond_th}
\displaystyle\frac{2(K^{\theta\theta}_{j})\delta t}{v_j^2 \delta\theta_k^2} \le 1.
\end{equation}
\end{itemize}
Note that the slowing-down coefficient $u$ as well as the diffusion tensor $K$ given in \eqref{eq:uK_def} depend only on $v$. 



The idea is then to use the explicit scheme (\ref{eq:scheme_st_1}) with the stability conditions (\ref{eq:stabcond_v}) and (\ref{eq:stabcond_th})  applied \textbf{locally} in each cell of the suprathermal grid. 
Indeed, the discrete scheme (\ref{eq:scheme_st_1}) corresponds to the finite volume formulation of a conservation equation where the time evolution of the $\alpha$ distribution function defined at the mesh centers is driven by the difference between the numerical fluxes calculated at the boundaries. The fluxes depend on the value of the distribution function in the neighboring cells. If the fluxes are applied during a time step $\Delta t$ which is too large with respect to the absolute values of the fields in the neighboring cells,  numerical instabilities occur. The idea is then to apply fluxes during a \textbf{limited} time step $\Delta t'$, possibly smaller than the imposed time step $\Delta t$. The time interval $\Delta t'$ is chosen such that the variation of the fields in the neighboring cells remain below their initial absolute values. Fluxes and fields are updated consistently at the frequency $\frac{1}{\Delta t'}$, until the imposed time step $\Delta t$ is reached.

\subsubsection{Stability and positivity}\label{sec442}

These conditions impose the stability of the explicit scheme \eqref{eq:scheme_st_1}, but not necessarily its positivity. Indeed, we have noticed that applying the explicit scheme \eqref{eq:scheme_st_1} with the stability conditions \eqref{eq:stabcond_v} and \eqref{eq:stabcond_th}  may lead to negative values of $f_\alpha^{ST}$ and thus lead to the development of numerical instabilities. This is especially true in the velocity region where the slowing-down coefficient $u$ is large, which may occur for example in the suprathermal region where $\alpha$-particles are created.   

A possible remedy is to introduce an "adaptative de-centering" in the discretization of the radial slowing-down current. We then go back to Eq.\,\eqref{eq:Jv_kj_discr1} and introduce the parameters $\eta_j$ such as:


\begin{equation}
\label{eq:Jv_kj_discr_eta}
J^v_{kj+1/2} = \frac12 u^v_{kj+1/2} \left[(1-\eta_j)f^n_{kj+1}+(1+\eta_j)f^n_{kj}\right]
-\frac{K^{vv}_{kj+1/2}}{\delta v_{j+1/2}}(f^n_{kj+1}-f^n_{kj}).
\end{equation}
The choice $\eta_j=0$ leads to the centered scheme \eqref{eq:Jv_kj_discr1}, while $\eta_j = 1$ leads to a pure upwind scheme. 
The decentering defined in \eqref{eq:Jv_kj_discr_eta} may also be seen as a perturbation of the discretized diffusion term.
Indeed, Eq.\,\ref{eq:Jv_kj_discr_eta} can be written in the following form:
\begin{equation}
\label{eq:Jv_kj_discr_eta2}
J^v_{kj+1/2} = \frac12 u^v_{kj+1/2} (f^n_{kj+1}+f^n_{kj}) -\widetilde{K}^{vv} \frac{f^n_{kj+1}-f^n_{kj}}{\delta v_{j+1/2}}.
\end{equation}
The stability condition \eqref{eq:stabcond_v} applied with the modified coefficient diffusion $\widetilde{K}^{vv} = K^{vv}_{kj+1/2}+\frac12  u^v_{kj+1/2}\eta_j\delta v_{j+1/2}$ instead of the original $K^{vv}$ defined in  \eqref{eq:uK_def} leads to the stability condition:
\[
\frac12 |u^v_{kj+1/2}|^2\delta t  \le K^{vv}_{kj+1/2}+\frac12 u^v_{kj+1/2}\eta_j\delta v_{j+1/2} \quad\mbox{and}\quad \frac{\delta t}{\delta v_{j+1/2}^2}\left(2 K^{vv}_{kj+1/2}+u^v_{kj+1/2}\eta_j\delta v_{j+1/2}\right)\le 1.
\]
Besides the positivity condition written in the case of an initial field $f_\alpha^{ST}$ localized in one velocity cell leads to:
\[
K^{vv}_{kj+1/2}+\frac12 u^v_{kj+1/2}\eta_j\delta v_{j+1/2}\ge 0 \quad\mbox{and}\quad \frac{1}{\delta v_{j+1/2}}\left(2 K^{vv}_{kj+1/2}+\frac12 u^v_{kj+1/2}\eta_j\delta v_{j+1/2}\right)\ge |u^v_{kj+1/2}|.
\]
The minimal value of $u_v\eta$ ensuring positivity is thus:
\begin{equation}
\label{eq:mineta}
u^v_{kj+1/2}\eta_j = \max\left\{0, |u^v_{kj+1/2}|-2K^{vv}_{kj+1/2}/\delta v_{j+1/2}\right\}.
\end{equation}
To ensure stability as well as positivity, we calculate the radial flux with respect to \eqref{eq:Jv_kj_discr_eta2} with $\eta_j$ given  by \eqref{eq:mineta} in each velocity cell. Actually, this amounts to using the scheme \eqref{eq:scheme_st_1} with the radial diffusion coefficient $K^{vv}$ replaced by:
\begin{equation}
\label{eq:Kvvmod}
\widetilde{K}^{vv}=\max\{K^{vv},|u^v|\delta v/2\}
\end{equation} 
and apply the conditions \eqref{eq:stabcond_v}. Note that in \eqref{eq:stabcond_v},  the condition imposed on the slowing-down coefficient $|u_{v}|\delta t\le\delta v$ is automatically fulfilled as soon as the one imposed on the (modified) diffusion coefficient $\widetilde{K}^{vv}$ is satisfied.

\subsection{Applying the stability condition locally}\label{sec45}

We describe now the accurate implementation of the algorithm, named \emph{Locally Sub-cycled Explicit} \textsc{LSE} algorithm that solves the problem of collisional relaxation of $\alpha$-suprathermal particles. The idea is to apply the explicit scheme \eqref{eq:scheme_st_1} with  the stability conditions \eqref{eq:stabcond_v} and \eqref{eq:stabcond_th}  applied \textit{locally} in each cell of the suprathermal grid.

Knowing the values of the distribution function $f_{jk}^n$ in any cell of the suprathermal velocity at time $t=t_n$, we  apply the following strategy:

{\bfseries First step} -- \emph{Local time steps calculation}\\
For each cell $\delta V_{jk}$ of the suprathermal velocity grid, we calculate a \textit{local} time step $\Delta t_{jk}$ such that the stability conditions in the $\theta$ and $v$ directions \eqref{eq:stabcond_th} are fulfilled. To find $\Delta t_{jk}$, the global time step, namely $\Delta t$, is halved until \eqref{eq:stabcond_th} and \eqref{eq:stabcond_v} are satisfied. The local time step $\Delta t_{jk}$ is then:
\begin{equation}
\Delta t_{jk}=\min(\Delta t_{jk}^\theta,\Delta t_{jk}^v),
\end{equation}
where:
\begin{equation}
\label{eq:local_dt_th}
\Delta t_{jk}^\theta = 2^{-{\rm nsplit}^{\theta}_{jk}} \Delta t ,
\end{equation}
and
\begin{equation}
\label{eq:local_dt_v}
\Delta t_{jk}^v = 2^{-{\rm nsplit}^{v}_{jk}}  \Delta t,
\end{equation}
${\rm nsplit}^{\theta}_{jk}$ (resp. ${\rm nsplit}^{v}_{jk}$) is the number of times the global time step has to be halved to fulfill the stability condition in the $\theta$ (resp. $v$) direction.

{\bfseries Second step} -- \emph{Sorting the cells}\\
Then, the cells of the suprathermal velocity grid are sorted with respect to their local time step $\Delta t_{jk}$ calculated above. This can for instance be done with an efficient algorithm (e.\ g., 'Heapsort' \cite{PRE924}), which takes on the order of $N\ln N$ operations for each time step where $N$ is the number of cells of the suprathermal velocity grid. This sorting stage then allows cells to be visited by the algorithm only when they actually need to be updated, and is thus an essential step for an computationally efficient algorithm, as shown in ref.\cite{LAR07}.

{\bfseries Third step} -- \emph{Sub-cycling}\\
Each cell has to be advanced in both directions $v$ and $\theta$ over a time $\Delta t$ with respect to its \textit{local} time-step $\Delta t_{jk}$, this procedure ensuring stability.
We thus have to perform a \textit{sub-cycling} for each cell. The effective computation proceeds through a loop over the smallest local time-step. Inside the loop, the fields (evaluated at the center of the cell) and the flux (evaluated at the borders) are updated consistently with the local time step of the considered cell. More precisely, we perform the following iterations:
\begin{equation}
\label{eq:scheme_st_1_sub_th}
\frac{f_{kj}^{p+1}-f_{kj}^p}{\Delta t_{jk}}  =  \frac{3v_j\delta v_j}{2\delta v^3_j}\frac{\sin\theta_{k+1/2}J^{\theta p}_{k+1/2j}-\sin\theta_{k-1/2}J^{\theta p}_{k-1/2j}}{\delta\mu_k}+\frac{1}{v_j^2}\frac{v_{j+\frac{1}{2}}^2J^{vp}_{kj+\frac{1}{2}}-v_{j-1/2}^2J^{vp}_{kj-1/2}}{2\delta v^3_j},
\end{equation}
where the superscript $p$ refers to the sub-cycled iterations. The sub-cycling starts with $f_{kj}^{p=0} = f_{kj}^n$ and ends after $p^{\max}_{jk}$ iterations where $\Delta t = p^{\max}_{jk} \Delta t_{jk}$. During the process, the flux $J^\theta_{k+1/2j}$ (resp. $J^v_{kj+1/2}$) defined in \eqref{eq:Jth_kj_discr1} (resp.\eqref{eq:Jv_kj_discr_eta2} and \eqref{eq:mineta}), are updated with a frequency corresponding to $1/\Delta t_{jk}^{\theta}$ (resp. $1/\Delta t_{jk}^{v}$). For more details on the sub-cycling method, we refer to \cite{LAR07}. This strategy guarantees stability and positivity everywhere on the suprathermal velocity grid.

By applying the local sub-cycling described above, we are able to treat the collisional part of the Vlasov-Fokker-Planck equation governing the suprathermal component of the $\alpha$ distribution function using a tractable explicit approach that does not lead to prohibitive computational time.

\begin{figure}
\centering
{\includegraphics[width=\textwidth]{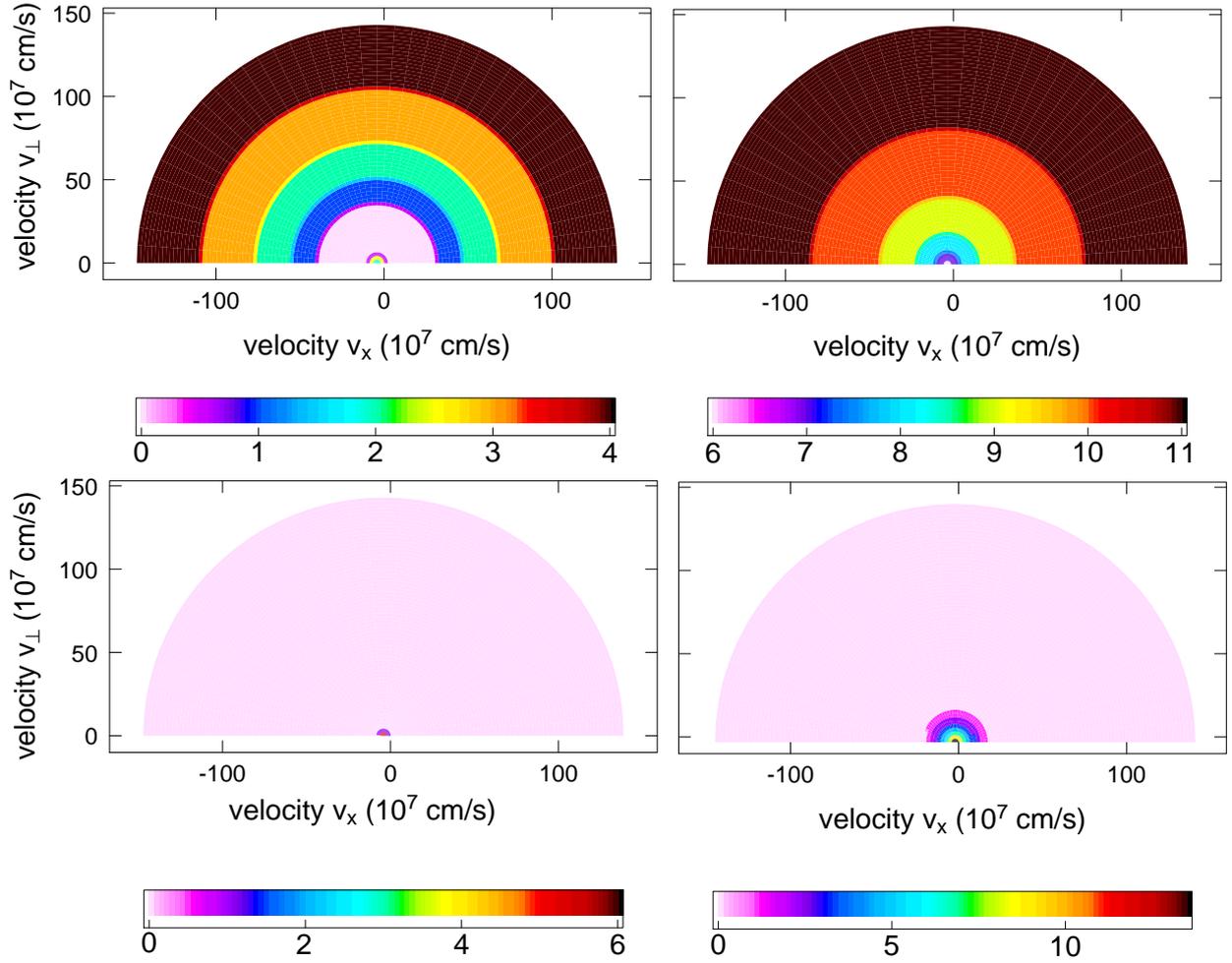}}
\caption{\label{fig:mapnsplit} Map of $nsplit^v$ (top) and $nsplit^{\theta}$ (bottom) represented in the suprathermal velocity grid in 2 locations. On the left, we consider a point in the hot spot where: $n_e \sim 10^{21}$ cm$^{-3}$ and $T_i\sim T_e \sim 0.5$ keV. On the right, we focus on a point taken in the dense shell where: $n_e \sim 10^{24}$ cm$^{-3}$ and $T_i\sim T_e \sim 0.01$ keV. Those conditions correspond to a typical implosion 1 ns before stagnation. Illustrations are given for a global time step $\Delta t=0.1$ ps
}
\end{figure}

To illustrate the efficiency of the \textsc{LSE} algorithm, we present in figure \ref{fig:mapnsplit}  the map of ${\rm nsplit}^{\theta}_{jk}$ and ${\rm nsplit}^{v}_{jk}$ defined in \eqref{eq:local_dt_th} and \eqref{eq:local_dt_v} on the suprathermal velocity grid. We consider two locations corresponding to the hot spot and the dense shell of a typical imploding capsule taken 1\,ns before stagnation. We note that the sub-cycling is more expensive in the dense shell region than in the hot spot. Indeed, the high density and low temperature of the shell imply smaller time step.

Furthermore, considering the maps of ${\rm nsplit}^{\theta}_{jk}$ represented at the bottom of figure\,\ref{fig:mapnsplit}, we note that to advance the fields in $\theta$, we mainly have to sub-cycle the most central cells, where the local time step imposed by the stability condition is the smallest since the local cell size $v_j \delta \theta$ is small close to the center. For the outermost velocity cells, no sub-cycling is actually needed. 



\subsection{Coupling with the thermal component}\label{sec46}

We now discuss the implementation of the coupling strategy between the suprathermal and the thermal components, as described by system
\eqref{eq:system_ST_T} in Sec.\,\ref{sec32}. 

\subsubsection{From the suprathermal point of view}\label{sec461}

\begin{figure}
\centering
{\includegraphics[width=0.5\textwidth]{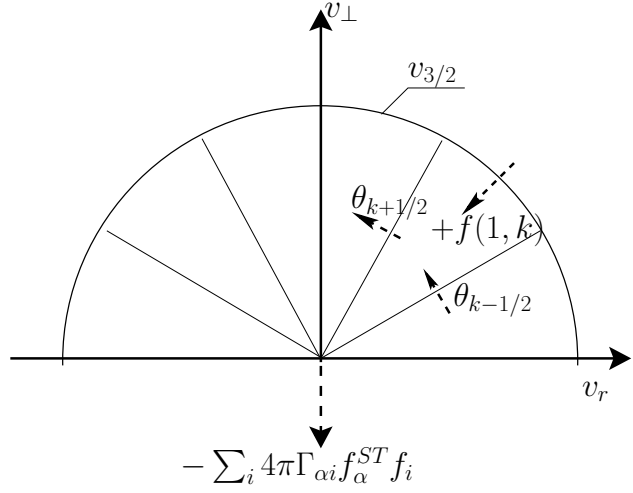}}
\vspace{-2cm}
\caption{\label{fig:polarcentre} Central mesh of the suprathermal velocity grid}
\end{figure}

From the point of view of suprathermal $\alpha$-particles, the coupling with the thermal component is made by the third term in Eq.\,\eqref{eq:third_term} on the right-hand side of \eqref{eq:FP_supra_discr}. It induces a time variation of the suprathermal distribution  given by the following equation:
\begin{equation}
\label{eq:coupl_st}
\left. \frac{\partial f^{ST}_\alpha}{\partial t}\right|_{ST \to T} = -\sum_{i} 4\pi\widetilde{\Gamma}_{\alpha i}\frac{A_\alpha}{A_i} f^{ST}_\alpha f_i \simeq -\sum_{i} 4\pi\widetilde{\Gamma}_{\alpha i}\frac{A_\alpha}{A_i} f^{ST}_\alpha n_i \delta^3(\vec v).
\end{equation}
The time evolution of the suprathemal distribution function in central velocity meshes is then governed by:
\begin{equation}
\label{eq:coupl_st_full}
\left.\frac{\partial f_{\alpha}^{st}}{\partial t}\right|_{\mbox{coll}} = \frac{1}{v^2}\frac{\partial}{\partial v}\left(v^2 J^v\right) + \frac{1}{v\sin\theta}\frac{\partial}{\partial \theta}\left(\sin\theta \,J^\theta\right)-\sum_{i} \widetilde{\Gamma}_{\alpha i}\frac{A_\alpha}{A_i} f^{ST}_\alpha n_i \displaystyle\frac{\delta(v)}{v^2},
\end{equation}
where the slowing-down currents $J^v$ and $J^{\theta}$ are given by Eq.(\ref{eq:Jv_st_discr}) and Eq.(\ref{eq:Jth_st_discr}) respectively.
As slowed down $\alpha$-particles approach the thermal velocity region, the transverse diffusion current $J^\theta$ intensifies so that the distribution function is almost isotropic in the central velocity meshes. Eq.(\ref{eq:coupl_st_full}) simplifies to:
\begin{equation}
\label{eq:coupl_st_full2}
\left.\frac{\partial f_{\alpha}^{st}}{\partial t}\right|_{\mbox{coll}} = \frac{1}{v^2}\frac{\partial}{\partial v}\left(v^2 J^v\right) -\sum_{i} \widetilde{\Gamma}_{\alpha i}\frac{A_\alpha}{A_i} f^{ST}_\alpha n_i \displaystyle\frac{\delta(v)}{v^2},
\end{equation}
where the slowing-down current $J_v$ can be approximated by:
$$
J^v \simeq \widetilde{\Gamma}_{\alpha i}\frac{A_\alpha}{A_i} \displaystyle\frac{n_i}{v^2} f^{ST}_\alpha.
$$
We then integrate Eq.\eqref{eq:coupl_st_full2} over a central mesh ($j=1$, $1\leq k \leq k_{\max}$) of the suprathermal velocity.
The suprathermal component in the central meshes corresponding to $j=1$ are then calculated as follows (see Fig.\ref{fig:polarcentre}):
\begin{equation}
\label{eq:coupl_st_discr}
\frac{f_{k1}^{n+1}-f_{k1}^n}{\Delta t} \frac{v_{3/2}^3}{3} = \sum_{i} n_i \widetilde{\Gamma}_{\alpha i} (f_{k3/2}^n-f_{k1}^n).
\end{equation}
In such a way, the distribution function remains stable in the most central part of the suprathermal velocity grid. 


\subsubsection{From the thermal point of view}\label{sec462}

To recover the full Fokker-Planck equation on the physical $\alpha$ distribution function $f_\alpha = f_\alpha^T + f_\alpha^{ST}$, we define an $\alpha$ thermal component $f_\alpha^T$, which evolves on the thermal velocity grid defined above. This is also the grid on which the thermal ion $D,T$ distribution functions evolve. This grid is actually inherited from the code \textsc{FPion}, so that we use the same cylindrical parametrization as explained in \cite{LAR93} for the $\alpha$ thermal component: $f_\alpha^T(r,v_r,v_\bot)$, $v_r$ and $v_\bot$ being the radial and tangential components of the velocity, respectively. 

The term \eqref{eq:third_term} subtracted from the suprathermal component equation reappears as a \textit{source term} in the Vlasov-Fokker-Planck equation governing the thermal component of the $\alpha$ distribution function $f_\alpha^T$, so that the relaxed suprathermal component feeds the thermal one and no $\alpha$ particle is lost in the process:
\begin{eqnarray}
&& \frac{\partial f_\alpha^T}{\partial t} + v_r\frac{\partial f_\alpha^T}{\partial r} +
\frac{v_\bot}{r}\left(v_\bot\frac{\partial f_\alpha^T}{\partial v_r}
- v_r\frac{\partial f_\alpha^T}{\partial v_\bot}\right) +
\frac{\mathcal{E}_\alpha}{A_\alpha}\frac{\partial f_\alpha^T}{\partial v_r} = \sum_{i} 4\pi\widetilde{\Gamma}_{\alpha i} \displaystyle\frac{\partial}{\partial \vec v} \cdot \,\left(\frac{A_\alpha}{A_i} f_\alpha^T \frac{\partial\mathcal{S}_i}{\partial \vec v}  - \nabla^2\mathcal{T}_i\frac{\partial f_\alpha^{T}}{\partial\vec v} \right) \nonumber \\
&& \qquad\qquad + \frac{1}{\widetilde{\tau}_{e\alpha}}\displaystyle\frac{\partial}{\partial \vec v} \cdot \,\left((\vec v - \vec {u_e})f_\alpha^T + \frac{T_e}{A_\alpha}\frac{\partial}{\partial \vec v}f_\alpha^T\right) +\sum_{i} 4\pi\widetilde{\Gamma}_{\alpha i}\frac{A_\alpha}{A_i} f^{ST}_\alpha f_i .
\label{eq:eqFP_t}
\end{eqnarray}
The source term coming from the slowing down of the suprathermal component appears in the last term on the right-hand side of \eqref{eq:eqFP_t}. From the point of view of the thermal component, the suprathermal component $f^{ST}_\alpha$ appears relatively constant over the whole thermal velocity grid since it varies significantly on the coarse suprathermal velocity grid whose mesh size is of the order of the thermal velocity. That is why we use the following estimate:
\begin{equation}
\label{eq:source_term_T}
\sum_{i} 4\pi\widetilde{\Gamma}_{\alpha i}\frac{A_\alpha}{A_i} f^{ST}_\alpha f_i \sim f^{ST}_\alpha (V_0)\sum_{i} 4\pi\widetilde{\Gamma}_{\alpha i}\frac{A_\alpha}{A_i}  f_i,
\end{equation}

$V_0$ being the mean ion velocity. This procedures guarantees an exact mass conservation: the number of particles that are removed form the suprathermal component are injected into the thermal component. Note that the source term feeding the $\alpha$ thermal component depends on the thermal distribution functions of \textbf{all} thermal ion species. To solve \eqref{eq:eqFP_t}, we use algorithms inherited from the code \textsc{FPion}. Their numerical implementation are for example discussed in \cite{LAR93}.

\subsection{Transport and acceleration of the suprathermal component}\label{sec47}

We discuss in this section the algorithm developed to solve the Vlasov part of  Eq.\,\eqref{eq:FP_supra_discr}), namely:
\begin{equation}
\label{eq:vlasov_st}
\frac{\partial f^{ST}_\alpha}{\partial t} + \vec v\cdot\vec\nabla_r f^{ST}_\alpha + \frac{ 
\vec {\mathcal E}_{\alpha}}{A_\alpha}\cdot\frac{\partial}{\partial \vec v} f^{ST}_\alpha = 0
\end{equation}
We deal with the advection and acceleration separately.

\subsubsection{Advection}\label{sec471}

In this stage, we solve the pure advection equation on the suprathermal component $f^{ST}_\alpha$ for a given velocity $\vec v$:

\begin{equation}
\label{eq:advec_st}
\frac{\partial f^{ST}_\alpha}{\partial t} + \vec v\cdot\vec\nabla_r f^{ST}_\alpha = 0,
\end{equation}
whose exact solution is given by:
\begin{equation}
\label{eq:soluce_advec}
f_\alpha^{ST}(\vec r,\vec v, t+\Delta t) = f_\alpha^{ST}(\vec r-\vec v \Delta t,\vec v, t).
\end{equation}
Thus, solving \eqref{eq:advec_st} amounts to interpolating \eqref{eq:soluce_advec} on the whole phase space. We thus start with a given point $(r,v,\theta)$ of the phase space, $v,\theta$ being chosen on the polar suprathermal velocity grid.
We have to compute the transformation of the suprathermal phase space coordinates $r,v,\theta$ during one time step $\Delta t$.
Since the suprathermal velocity grid is centered on the mean bulk velocity $V_0$, we firstly project the polar velocity coordinates on the cylindrical basis:
\begin{equation}
\label{eq:proj1}
v_r=V_0+v\cos\theta,  \qquad v_\bot = v\sin\theta.
\end{equation}  
Then, we apply the following transformations on $r,v_r,v_\bot$ over one time step $\Delta t$:
\begin{equation}
\label{advvr} 
 r(t-\Delta t) = \left[r(t)^2-2r(t)v_r(t)\Delta t+ v^2\Delta t^2 \right]^{1/2},\quad
v_r(t-\Delta t) = \frac{r(t)v_r(t)- v^2\Delta t}{r(t-\Delta t)},\quad v_\bot(t-\Delta t) = \frac{r(t)v_\bot(t)}{r(t-\Delta t)},
\end{equation}
which gives us the advected point in phase space. For the interpolation in space, we have to find the two consecutive nodes $r_{i_0}$ and $r_{i_0+1}$ of the spatial mesh such that $r_{i_0} \leq r(t-\Delta t) \leq r_{i_0+1}$. Then, for each spatial nod $r_{i_0}$ (respectively $r_{i_0+1}$), we have to carry out an interpolation of \eqref{advvr} on the polar suprathermal velocity grid centered on the local mean bulk velocity $V_0(r_{i_0})$ (respectively $V_0(r_{i_0+1})$).
We thus calculate:
\begin{equation}\label{advv}
 v(t-\Delta t) = \left[\left(v_r(t-\Delta t)-V_0(r_i) \right)^2+v_\bot^2(t-\Delta t)\right]^{1/2} , \quad \theta(t-\Delta t) = \cos^{-1}{\frac{v_r(t-\Delta t)}{v(t-\Delta t)}},
\end{equation}
for $i=i_0$ and $i=i_0+1$. We then interpolate \eqref{advv} on the nodes of the suprathermal velocity grid centered on $V_0(r_i)$, using a simple linear interpolation method. This gives us the advected points:
\begin{equation}
f_{i_0}=f_\alpha^{ST}(r_{i_0},v(t-\Delta t),\theta(t-\Delta t),t-\Delta t),\quad 
f_{i_0+1}=f_\alpha^{ST}(r_{i_0+1},v(t-\Delta t),\theta(t-\Delta t),t-\Delta t).
\end{equation}
The final stage is a cubic interpolation with respect to space:
$$ f_\alpha^{ST}(r(t-\Delta t),v(t-\Delta t),\theta(t-\Delta t),t-\Delta t) = f_{i_0}+p\delta rf_{i_0}^\prime
   + p^2[3\delta f-\delta r(2f_{i_0}^\prime+f_{i_0+1}^\prime)]    + p^3[\delta x(f_{i_0}^\prime+f_{i_0+1}^\prime)-2\delta f] $$
with $ \delta r = r_{i_0+1} - r_{i_0}, \quad p=r(t-\Delta t)/\delta r, \quad \delta f = f_{i_0+1}-f_{i_0}$. In this equation, the spatial gradients $f_{i_0}^\prime$ and $f_{i_0+1}^\prime$ are evaluated by finite differences. The slopes are limited to prevent unphysical over/undershoots in the interpolation process.


\subsubsection{Acceleration}\label{sec472}

The electric field effect on the $\alpha$ suprathermal component is modeled by:
\begin{equation}\label{accf}
\frac{\partial f_\alpha^{ST}}{\partial t} + \frac{\vec{\mathcal E}_\alpha}{A_\alpha}
\frac{\partial f_\alpha^{ST}}{\partial \vec v} = 0
\end{equation}
where the effective electrostatic field $\vec {\mathcal E}_\alpha$ is defined by Eq.\,\eqref{Ei}. Here again, we use a method of characteristics to solve \eqref{accf} since an acceleration can be seen as an advection in velocity.  The situation gets simpler here, since we only have to carry out an interpolation in velocity on the suprathermal velocity grid. The process is repeated independently in each spatial cell.

\subsection{Chain of algorithms to solve the suprathermal Vlasov-Fokker-Planck problem}\label{sec48}

We conclude this section by summarizing the sequence of algorithms that have been developed to solve the whole problem of creation, transport and collisional relaxation of $\alpha$ suprathermal particles, consistently with a ion-kinetic treatment of the plasma thermal bulk. In particular, we show how the algorithms related to the suprathermal components are linked with those dealing with electrons and thermal ion distribution functions.
This constitutes the main loop of our kinetic code \textsc{Fuse}. For a global time step $\Delta t$, we apply the following splitting sequence:

{\bfseries Step 1} -- \emph{Electron conductivity}\\
We solve the conduction part of \eqref{eqTe}, which takes the form of  a pure diffusion (or heat) equation during the time $\Delta t/2$.

{\bfseries Step 2} -- \emph{Acceleration}\\
We accelerate ion thermal distribution functions for species D, T, $\alpha$ over the time $\Delta t/2$, and at the same time we solve the convective part of \eqref{eqTe}, which enables us to improve the energy conservation between ions and electrons (see \cite{LAR03A}).  Then, we accelerate the suprathermal $\alpha$  component.

{\bfseries Step 3} -- \emph{Advection}\\
We carry out the advection of thermal components for every ion species D, T, $\alpha$ as well as the suprathermal $\alpha$  component over the time $\Delta t/2$.


{\bfseries Step 4} -- \emph{Feeding the suprathermal component}\\
The suprathermal $\alpha$  component is fed by the fusion reaction according to \eqref{eq:evol_fdh_source} applied over the whole time step $\Delta t$.

{\bfseries Step 5} -- \emph{Suprathermal collisional relaxation}\\
We next solve the collisional part of \eqref{eq:FP_supra_discr} applying the Locally Split Explicit (LSE) algorithm over the time step $\Delta t$.

{\bfseries Step 6} -- \emph{Feeding the thermal component}\\
We apply the feeding term \eqref{eq:source_term_T} of the $\alpha$ thermal component by the suprathermal one over the time step $\Delta t$.

{\bfseries Step 7} -- \emph{Thermal collisional relaxation}\\
We perform the collisional relaxation of every ion thermal distribution functions (for ion species D, T, $\alpha$) on thermal ions and on electrons, applying the same algorithms as in \textsc{Fpion}. Note that the collisional relaxation of ion distribution functions on themselves is non-linear and is solved using Crank-Nicholson iterations with an ADI scheme (see Appendix of \cite{CHE979}).

{\bfseries Step 8} -- \emph{Advection}\\
Step 3 is repeated for another $\Delta t/2$.

{\bfseries Step 9} -- \emph{Acceleration}\\
Step 2 is repeated for another $\Delta t/2$.

{\bfseries Step 10} -- \emph{Electron conduction}\\
Step 1 is repeated for another $\Delta t/2$.

After each modification of the ion distribution functions (thermal or suprathermal), the ion moments as well as the slowing-down and diffusion coefficients are updated consistently.

\subsection{Validation of the code by test problems}\label{sec49}

In this section, we apply the algorithms developed to model the collisional relaxation and thermalization of $\alpha$-particles in simplified configurations where analytical results are known. 

\subsubsection{Isotropic time-dependent test problem}\label{sec491}

In this first test problem, we consider the collisional relaxation of fast $\alpha$-particles in an homogeneous and steady plasma made of one mean ion species $Z_i=1,A_i=2.5$ and electrons. The reference density is $n_i=n_e=10^{22}$ particles/cm$^3$, and the temperature is 1 keV. We keep those conditions  constant during the test problem calculation. Suprathermal $\alpha$ particles are then injected isotropically at the energy 3.52 MeV at a steady rate $S_0$ (particles.cm$^{-3}$.s$^{-1}$), so that the suprathermal component remains isotropic during the slowing down process. Following our two-scale approach, the $\alpha$ distribution function $f_\alpha(v,t)=f_\alpha^{ST}(v,t)+f_\alpha^{T}(v_r,v_\bot,t)$ is the solution of:

\begin{eqnarray}
\label{eq:system_ST_T_tp1}
&& \left.\partial_t f_\alpha^{ST}\right.=  \Gamma_{\alpha i} \frac{n_i}{v^2} \partial_{v} f_\alpha^{ST} + \frac{1}{\tau_{\alpha e} v^2}\partial_v\left(v^3 f_\alpha^{ST}\right)- 4\pi n_i\Gamma_{\alpha i} f_\alpha^{ST} \frac{\delta(v)}{4\pi v^2}+\frac{{S_0}\delta(v-v_h)}{4\pi v^2}, \nonumber\\
&& \left.\partial_t f_\alpha^T\right. = \left.\partial_t f_\alpha^T\right|_{\alpha i} + \left.\partial_t f_\alpha^T\right|_{\alpha e}+ 4\pi\Gamma_{\alpha i}  f_i  f_\alpha^{ST}(0).
\end{eqnarray}

$\left.\partial_t f_\alpha^T\right|_{\alpha i}$ (resp. $\left.\partial_t f_\alpha^T\right|_{\alpha e}$) corresponds to the collisional terms of the thermal ions (resp. electrons) on the $\alpha$-thermal particles.

\begin{figure}
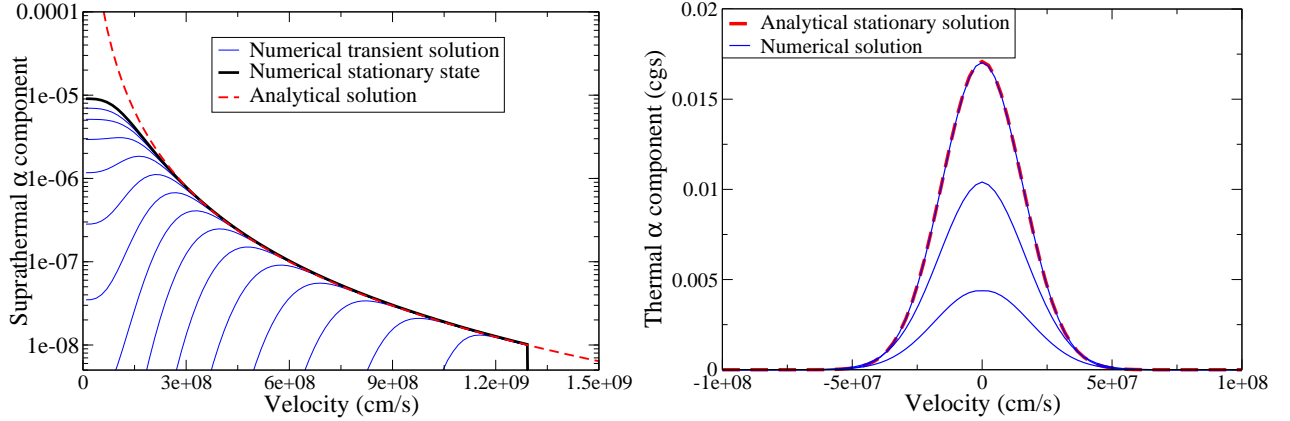

        \centering
        \begin{subfigure}[b]{0.5\textwidth}
                \rotatebox{0}{\includegraphics[width=\textwidth]{comp_fdhot_soluce1.eps}}
                \caption{Time evolution of the $\alpha$ suprathermal component for $t\leq\tau_s$. The chosen time interval between 2 consecutive curves is 0.05 ns. The exact solution is represented in dashed lines. }
                \label{fig:fst1}
        \end{subfigure}%
        ~ 
        \begin{subfigure}[b]{0.5\textwidth}
                \rotatebox{0}{\includegraphics[width=\textwidth]{fd_cold_sym.eps}}
                \caption{Time evolution of the $\alpha$ thermal component for $t\leq 4$ ns. The suprathermal source term is shut down after 1 ns, the $\alpha$ thermal component relaxes towards the Gaussian.}
                \label{fig:ft1}
        \end{subfigure}
        \caption{\label{ftp1}
Time evolution of the $\alpha$ distribution function corresponding to the isotropic test problem. Distribution functions are expressed in cgs units, namely in cm$^{-4}$.s$^{-1}$. The $\alpha$ source term is shut down after $\tau_s=1$ ns.}
\end{figure}

In those conditions, we have the characteristic velocity scales, expressed in cm/s:
\begin{equation}
v_i^{th} \sim 3.0\times 10^7 << v_c \sim 1.1\times 10^8 << v_h\sim 1.3\times 10^9 < v_e^{th} \sim 4.2\times 10^9 
\end{equation} 
For $v>v_{c}$ ($v_c$ given in Eq.\,\eqref{eq:vcoup}), the slowing down of $\alpha$-particles is mainly due to the Coulomb collisions with electrons. The suprathermal component $f_\alpha^{ST}(v,t)$ then tends to the stationary solution of:
\begin{equation}
\partial_t f_\alpha^{ST} = \frac{1}{\tau_{\alpha e} v^2}\partial_v\left(v^3 f_\alpha^{ST}\right)+\frac{{S_0}\delta(v-v_0)}{4\pi v^2}.
\end{equation}
The stationary solution is given by:
\begin{equation}
\label{eq:statio_tp1}
f_1(v) = \frac{S_0\tau_{\alpha e}}{v^3}\mathcal{H}(v_0-v), v>v_c,
\end{equation}
where $v_h$ is the velocity corresponding to the injected $\alpha$-particles at 3.52 MeV, which corresponds to $v_h \sim 1.3\times 10^9$ cm/s, and $\mathcal{H}$ is the Heaviside distribution.
We plot $f_\alpha^{ST}(v,t)$ calculated by \textsc{Fuse}  at different times as well the stationary analytical solution given by (\ref{eq:statio_tp1}) (see Fig (\ref{ftp1})). The numerical solution agrees with (\ref{eq:statio_tp1}) as long as $v>v_c$. When $v<v_c$, ions tend to dominate the slowing down of the $\alpha$-particles and the suprathermal component solution of (\ref{eq:system_ST_T_tp1}) tends to a stationary state that is almost constant close to thermal ions. This is due to the removal of the term $\propto f_\alpha^{ST} n_i\delta^3(\vec v)$ in the collision term governing the slowing down of $f_\alpha^{ST}$. The suprathermal component actually feeds the thermal one, the feeding process being driven by the source term $\propto f_\alpha^{ST}(v=0)f_i$. The thermal component subsequently evolves towards a Maxwellian characterized by the total density $n_\alpha$ of $\alpha$ particles injected in the system, and the reference temperature $T_0$ (which is kept constant during the test problem calculation):
\begin{equation}
\label{eq:gauss_tp1}
\mathcal{M}_\alpha(v)=n_\alpha \left(\frac{m_\alpha}{2\pi T_0}\right)^{3/2}\exp{-\frac{m_\alpha v^2}{2T_0}}.
\end{equation}
The total density is given by: 
\begin{equation}
n_\alpha = \int_0^{\tau_s} S_0 dt,
\end{equation}
$\tau_s$ being the time when the source is shut down. 
The convergence to the Gaussian (\ref{eq:gauss_tp1}) is represented on Fig (\ref{ftp1}). Note that this convergence is calculated on the refined thermal grid. The $\alpha$ thermal component is fed by a source term $\propto f_i$, of width $\sim\sqrt{T_0/m_i}$, and relaxes on the thermal grid towards the Gaussian \eqref{eq:gauss_tp1} of width $\sim\sqrt{T_0/m_\alpha}$.

\subsubsection{Anisotropic time-dependent test problem}\label{sec492}

\begin{figure}[h]
\centering
{\includegraphics[width=\textwidth]{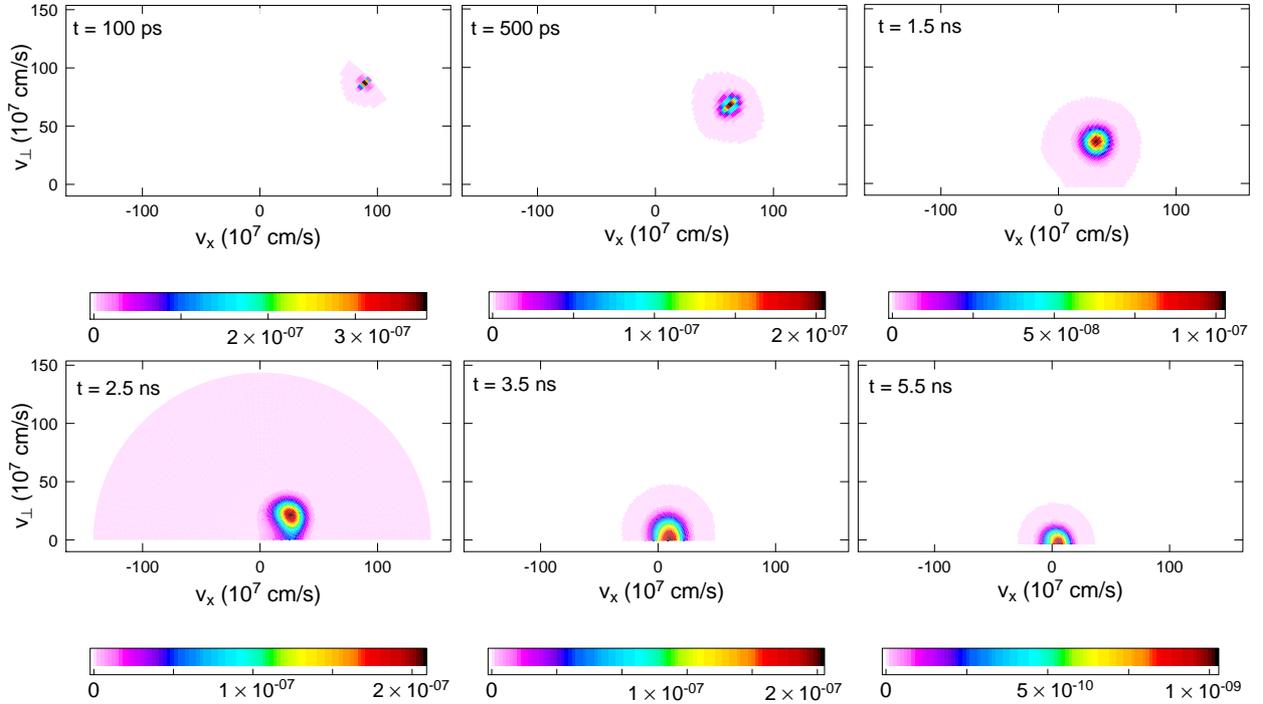}}
\caption{\label{tp_fdhot_2} $\alpha$ suprathermal distribution solution of the anisotropic test problem at different times. Final stages of collisional relaxation. The values of the distribution function are expressed in cgs units.
}
\end{figure}

\begin{figure}[!h]
\centering
{\includegraphics[scale=0.31]{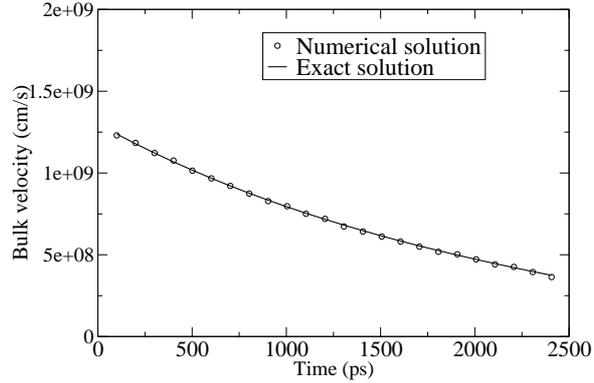}}
\caption{\label{vmax_2} Time evolution of the velocity corresponding to the maximum of the $\alpha$ suprathermal distribution function solution corresponding to the anisotropic test problem.
}
\end{figure}

We next consider the following anisotropic test problem. We consider an initial condition for the $\alpha$ suprathermal component highly localized in velocity space. Namely, we take:
\begin{equation}
f_\alpha^{ST}(v,\theta,t=0) = n_\alpha \frac{\delta(v-v_0)}{4\pi v^2}\delta(\cos\theta-\cos\theta_0),
\end{equation}
with $v_h=1.3\times10^9$cm/s and $\theta_0=\pi/4$. 
We then let the suprathermal $\alpha$ distribution slow dow on electrons and on thermal ions. 
As previously, the thermal plasma is homogeneous and made of one ion species $Z_i=1,A_i=2.5$ and electrons. The temperature of the thermal plasma is kept constant during the calculation: we take $T_0=5$ keV.
In those conditions, the characteristic velocity scales are (in cm/s):
\begin{equation}
v_{th,i} \sim 6.9\times 10^7 << v_c \sim 2.4\times 10^8 << v_h\sim 1.3\times 10^9 < v_{th,e} \sim 9.4\times 10^9 
\end{equation} 
The evolution of the $\alpha$ distribution function is represented in Fig.\ref{tp_fdhot_2}.
As long as $v>v_c$, the momentum and energy losses by the fast ions to the background plasma electrons are the dominant process. The distribution function remains highly localized in velocity space around a velocity $v_{\mbox{b}}(t)$ that declines due to the slowing down on electrons. The velocity of the bulk $v_{\mbox{b}}(t)$ can be calculated analytically \cite{RAX}:
\begin{equation}
\label{eq:bulk}
v_{\mbox{b}}(t)=\lbrack(v_0^3+v_c^3)\exp{-\frac{3t}{\tau_{\alpha e}}}-v_c^3\rbrack^{1/3}
\end{equation}
The comparison between the code and the exact solution is represented on Fig.\ref{vmax_2} and reveals a pretty good agreement, as long as $v>v_c$. Then, as $v\leq v_c$, the energy diffusion process as well as the perpendicular diffusion due to the thermal ions become significant.
The $\alpha$ distribution function is scattered in the $\theta$ direction, due to the diffusion on the thermal ions, that intensifies as $v\to 0$. Consequently, as $v\to 0$, the $\alpha$ suprathermal distribution tends to become isotropic while feeding the thermal component. Finally, the thermal component then converges towards the Gaussian, as in the first test problem. To model properly what happens in the vicinity of the thermalization, for $v\sim v_i^{th}$, we solve the full Coulomb operator applied to the $\alpha$ thermal component $f_\alpha^T$ that evolves on the thermal refined grid. This guarantees a proper modeling of the thermalization of the $\alpha$ distribution function, as it slows down, scatters and diffuses in energy in joining up with the background thermal ions.

\subsubsection{Energy conservation}\label{sec493}

\begin{figure}[!h]
\centering
{\includegraphics[scale=0.27]{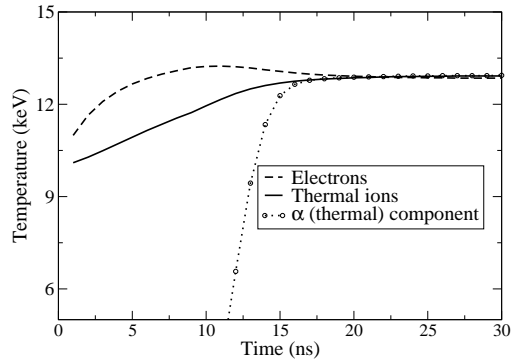}}
\caption{\label{massener} Time evolution of electron and thermal ion temperatures corresponding to a pure collisional relaxation test problem.
}
\end{figure}

We finally consider a full collision relaxation process, starting from an isotropic $\alpha$ suprathermal component that slows down through collisions on the electrons and the thermal ions. In this test problem, the electron (res. ion) temperatures evolve consistently with the slowing down of the suprathermal particles. More precisely, as $v>v_c$, suprathermal particles slow down essentially on electrons. The electron temperature thus increases. Then, due to the collisional relaxation of thermal ions with electrons, the thermal ion temperature increases. When the suprathermal particles reach the thermal velocity region, the $\alpha$ thermal component builds up and a collisional relaxation between electrons and thermal ions (including the $\alpha$ thermal component) brings the system to a stationary state. The aim of this test problem is to illustrate that the way we solve the coupling between the suprathermal component and the thermal background ensures the conservation of mass and energy. We check that the total mass remains constant (with a numerical error less than 1\% due to the finite size of the velocity mesh). We plot the time evolution of the temperatures (electrons, thermal background ions and $\alpha$-thermal component) on Fig.(\ref{massener}). We show how the system evolves naturally to a stationary state calculated by the algorithm described above. The total energy variation of the system between the initial state  and the final stationary state is less than 1\%.  
\vspace{0.4cm}

Our original algorithm based on a 2-scale approach to model the collisional relaxation between suprathermal particles and the thermal background is thus validated in simplified test problems where exact results are known. Besides, the mass and energy conservation principles are fulfilled at a discrete level. We can consider that our code \textsc{Fuse} is reliable. We then apply it on real target configurations.

\section{Application on the ignition and thermonuclear burn of typical ICF capsules}\label{sec5}

We apply the numerical scheme presented in Sec.\,\ref{sec4} to model a typical spherical implosion of a cryogenic DT capsule. Our code allows us to study ion-kinetic effects during  the ignition stage and the beginning of the thermonuclear burn stage.

\subsection{Initial conditions}\label{sec51}

We consider the same fluid reference simulation as in \cite{LAR03A} corresponding to an ICF target with parameters typical of ignition capsules designed for the LMJ and NIF laser \cite{SAI001} and \cite{BRA012}. Namely, we consider a 0.3\,mg  cryogenic DT layer deposited on the inner surface of a CH shell of a 1\,mm (inner) radius.  The kinetic calculation is started at $t=17$\,ns after the beginning of the implosion, when the main converging shock reaches the center of the target. The boundary condition is taken from the hydrodynamic simulation. The densities, temperatures and velocities are recorded on the fuel/pusher interface in the fluid simulation.

The kinetic simulation considers three ion species, namely D, T and $\alpha$. Initially, only thermal species D and T are present. They give birth to suprathermal $\alpha$  particles in the fusion reactions. The relaxation of the suprathermal $\alpha$  component then leads to the creation of an $\alpha$ thermal component interacting with the other thermal ion distribution functions (D and T, respectively). Note that the thermal bulk is described in more details than in \cite{LAR03A} where a single mean ion species with a mass number of 2.5 was considered. 

In our kinetic simulation, the position of each spatial meshes is updated after each time step with respect to the imposed boundary condition and to the fixed number of spatial meshes $i_{\max}$. This updating is  performed before each advection phase. This means that the position of a given spatial cell $r_{i_0}$, with $1\leq i_0 \leq i_{\max}$ is time dependent, decreasing with the size of the imploding system.  To represent in a satisfactory manner both the dense region where the fluid simulation grid is the finest and the central zone where it is rather coarse, we employ 78 cells with a geometrically varying mesh size (with the ratio 0.97) so that the mesh size $\delta r$ is decreasing from 20\,$\mu$m near the center to less than one micron near the outer boundary. The thermal velocity space $(v_r,v_\bot)$ is discretized into $129\times 64$ cells, whereas the suprathermal velocity grid $(v,\theta)$ makes use of $100\times60$ cells. The reference time-step value is 0.05\,ps.

\subsection{Comparison with \textsc{Fpion} and \textsc{FCI1}}

\begin{figure}[!h]
        \centering
   \includegraphics[width=0.65\textwidth, angle=0]{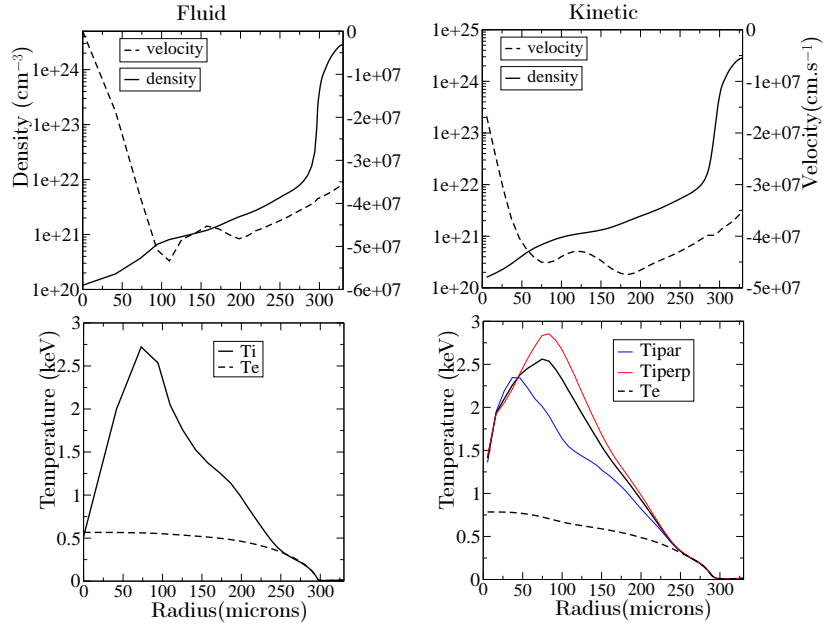}
   \caption{\label{hydr100}Profiles of the density, velocity and of the electron and total ion temperatures  in a DT ignition target  at the time $t=17.1$ ns, which corresponds to 100\,ps after the beginning of the kinetic calculation and roughly 1\,ns before the target stagnation.}
\end{figure}

\begin{figure}[!h]
        \centering
   \includegraphics[width=0.65\textwidth, angle=0]{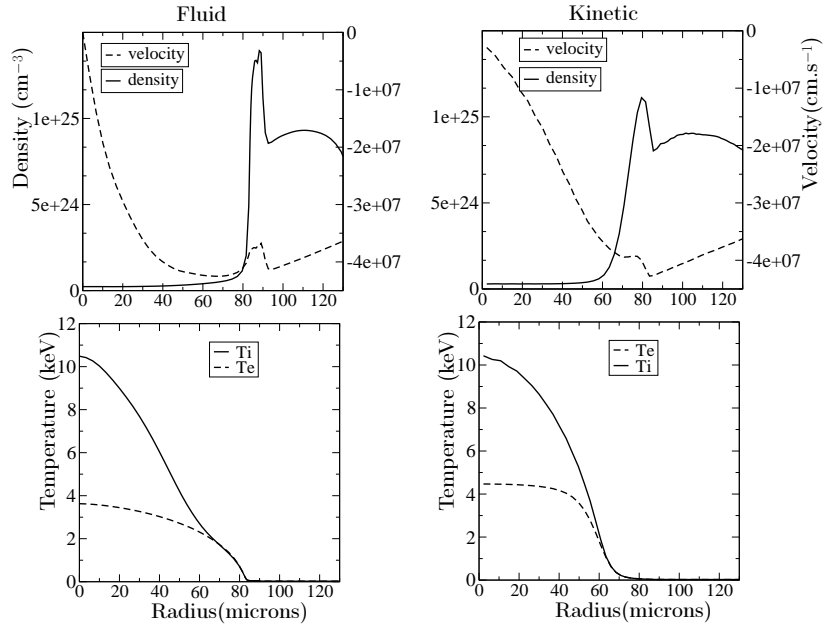}
   \caption{\label{hydr650}Profiles of the density, velocity and of the electron and total ion temperatures  in a DT ignition target  at the time $t=17.65$ ns, which corresponds to 650\,ps after the beginning of the kinetic calculation. This time is also just before the target stagnation.}
\end{figure}

To validate the thermal part of our code \textsc{Fuse}, we compare the density, velocity and temperature profiles with the hydrodynamic code \textsc{FCI1} as well as with the kinetic code \textsc{FPion} at two different times of the implosion:
\begin{itemize}
\item{at $t=17.1$ ns, that is to say 100 ps after the beginning of the implosion. We find a pretty good agreement between the \textsc{Fuse} kinetic calculation and the \textsc{FCI1} fluid simulation (Fig.\ref{hydr100}). The kinetic modeling reveals a significant anisotropy on the ion temperatures (and pressures), as the one observed with \textsc{FPion} \cite{LAR03A}. The anisotropy then tends to disappear during the implosion.}
\item{At $t=17.65$ ns, in the vicinity of the target stagnation, \textsc{Fuse} and \textsc{FCI1} are still in good agreement. However, we note that the compression zone near the inner interface of the dense fuel lies closer to the target center in the kinetic calculation (see the negative velocity gradient region about r = 70 $\mu$m on the right part of Fig.\ref{hydr650}). This result has already been obtained with \textsc{FPion} and discussed in \cite{LAR03A}. This is related to a higher ion heat flux, which tends to increase the rate of ablation of the cold fuel by the hot spot.  
}
\end{itemize}
As long as $t\leq 17.65$ ns, the $\alpha$-particles number is small, so that the above comparisons between the codes \textsc{Fuse} and \textsc{Fpion} (which does not take $\alpha$-particles into account) are relevant and tend to validate the methods programmed in  \textsc{Fuse}  regarding the thermal background (thermal ions and electrons).    

\subsection{Transport of $\alpha$ particles}\label{sec52}

\begin{figure}[!h]
        \centering
  \includegraphics[width=0.45\textwidth]{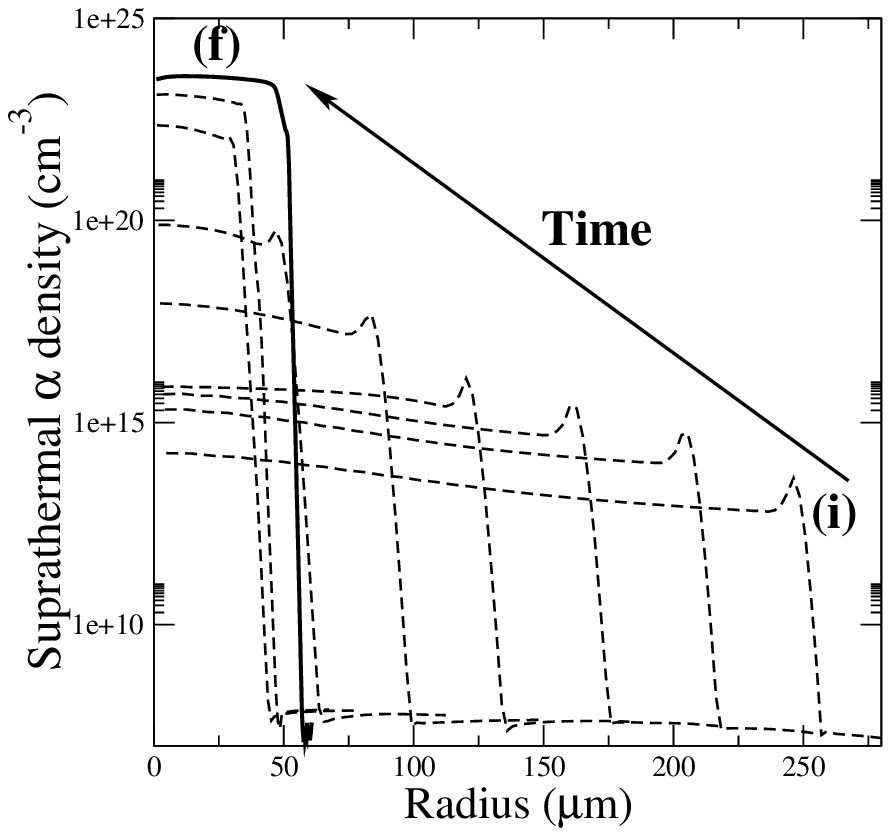}
  \includegraphics[width=0.45\textwidth]{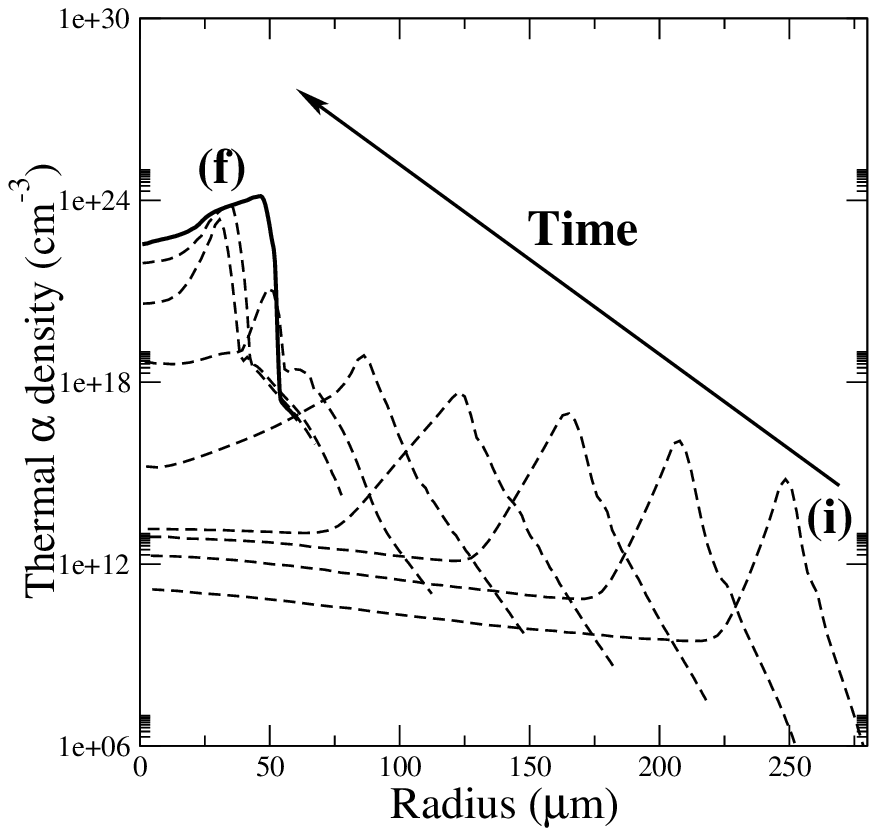}
  \caption{Density profiles of suprathermal (left) and thermal (right) $\alpha$ particles. The initial time (i) corresponds to $t=17.1$ ns and the final time (f) to $17.87$ ns. The time interval between two consecutive profiles is 50\,ps.}\label{fig:nacnah}
\end{figure}

We analyze the transport of suprathermal $\alpha$  particles throughout the capsule. Figure\,\ref{fig:nacnah} shows the spatial density  profiles during the implosion for the suprathermal and thermal  components of $\alpha$-particles. At early times, suprathermal $\alpha$-particles are produced in the hot central region of the capsule and deposit their energy in the surrounding cold shell. The region corresponding to the suprathermal $\alpha$ energy deposition is indicated by a sharp decreasing of the suprathermal density profile. This occurs at a distance which corresponds to the collisional mean free path of suprathermal $\alpha$  particles.  Meanwhile, the slowing down of suprathermal $\alpha$  particles feeds the thermal component, that process corresponding to the bump observed in the thermal $\alpha$ density profiles (Figure\,\ref{fig:nacnah}-right). 

During the implosion process, the $\alpha$ collisional mean free path decreases, so that the $\alpha$ suprathermal particles are trapped in a smaller radius. In the mean time, the production of suprathermal $\alpha$-particles  intensifies due to the increasing ion temperature. As a result, the suprathermal $\alpha$  density increases.   

\subsection{Collisional relaxation of suprathermal $\alpha$ particles}\label{sec53}
\subsubsection{Anisotropy in the suprathermal region}\label{sec531}

\begin{figure}[!h]
\centering
{\includegraphics[width=\textwidth]{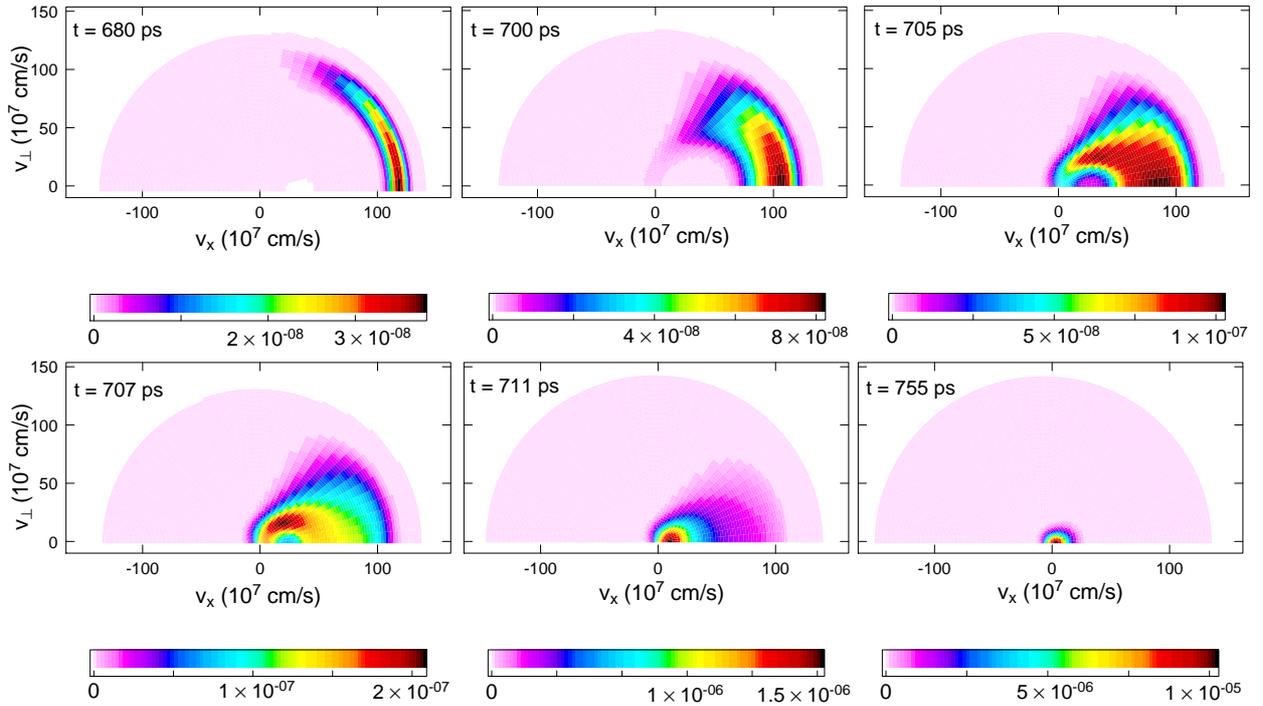}}
\caption{\label{fdhot_1} $\alpha$ suprathermal distribution observed in a given mesh of the imploding hot spot at different times. The simulation takes into account the creation, the transport and the collisional relaxation of $\alpha$ particles. The values of the distribution function are expressed in cgs units. Times refer to beginning of the kinetic calculation.
}
\end{figure}

In this section, we focus on the collisional relaxation of the suprathermal $\alpha$  component. We consider a given spatial cell with the number $i_0$ that evolves in space during implosion. The distribution function of $\alpha$-particles   $f_\alpha^{ST}(r_{i_0}(t),v,\theta,t)$ is presented in figure \ref{fdhot_1}.

The suprathermal distribution function is rather anisotropic. It is highly peaked toward positive velocities $v_r > 0$. This can be explained by the inhomogeneous fusion reaction source term, which strongly depends on the ion local temperature. Since $T_i$ is more peaked towards the center of the capsule, as it can be seen in the temperature profiles in figure\,\ref{hydr100}, an observer located outside of the highly emissive central region sees the suprathermal $\alpha$-particles passing from the center to the outside. That leads to a local distribution shape shown in the top panel of figure\,\ref{fdhot_1}. The spatial gradient of the fusion reaction source term \eqref{eq:evol_fdh_source} thus accounts for the anisotropy of the suprathermal $\alpha$ distribution function.

Let us consider the cell $i_0$ with the radius such that $r_{i_0}(t) = \lambda_\alpha(\rho(t))$, where $\lambda_\alpha$ is the collisional mean free path of a suprathermal $\alpha$ particle and $\rho$ the mean density of the capsule. As $\alpha$-particles deposit their energy in the considered spatial cell $i_0$, which corresponds to the sequence shown in figure\,\ref{fdhot_1}, the suprathermal $\alpha$ distribution function slows down significantly towards the thermal velocity region. During this slowing down process, the distribution function tends to spread over a wider domain in the polar angle $\theta$. This is a consequence of the diffusion part of the Fokker-Planck equation, which leads to a mainly transverse slowing-down current that intensifies close to the thermal velocity region. 

\begin{figure}[!h]
\centering
{\includegraphics[width=\textwidth]{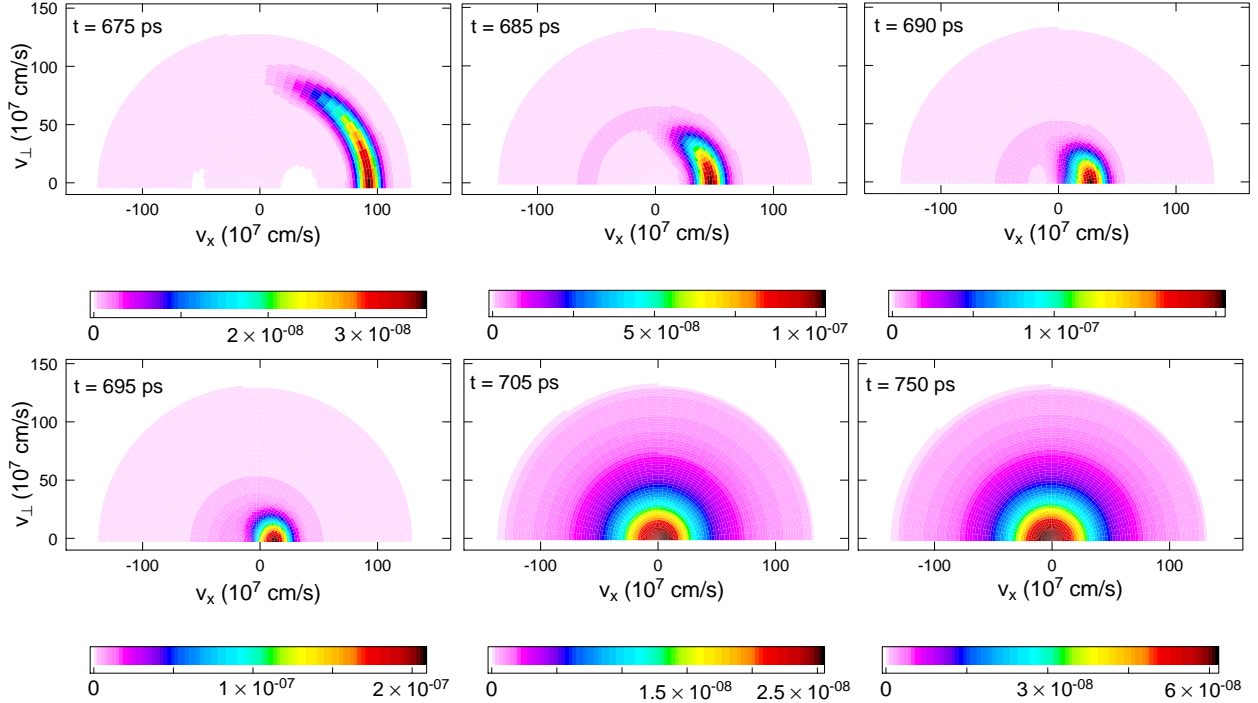}}
\caption{\label{fdhot_2} $\alpha$ suprathermal distribution observed in a given mesh of the imploding  hot spot at different times, when only the collisional relaxation is considered, starting from a given anisotropic initial state. Times refer to beginning of the kinetic calculation.
}
\end{figure}

To check that the collisional module of the code behaves correctly in a real target configuration, we artificially do not calculate the effect of the advection and acceleration on the $\alpha$-suprathermal component, so that the time evolution is driven by the collisions on  electrons and thermal ions only. The corresponding time evolution is represented in Fig.\ref{fdhot_2}. This numerical test is closed the third test problem presented in Sec.4.9.3, but is carried out in thermodynamic conditions corresponding to real ICF target configuration. The suprathermal particles are initially distributed anisotropically in velocity space with respect to Fig.\ref{fdhot_2} (top-left). For $v\geq v_c\sim 3-4 v_i^{th}$, fast ions mostly slow down by collisional drag on the background electrons with very little pitch-angle scattering. The fast ions stay mostly in their original pitch-angle direction. For $v\leq v_c$, the suprathermal particles slow-down predominantly on the thermal background ions and scatter in pitch-angle. The suprathermal distribution function tends to be isotropic as it approaches the thermal velocity region. The suprathermal grid resolution is fine enough to represent the variations of the suprathermal component, that tends to be constant as it gets closer to the thermal velocity region.

\subsubsection{Feeding the thermal component }\label{sec532}

\begin{figure}[!h]
\centering
{\includegraphics[width=0.9\textwidth]{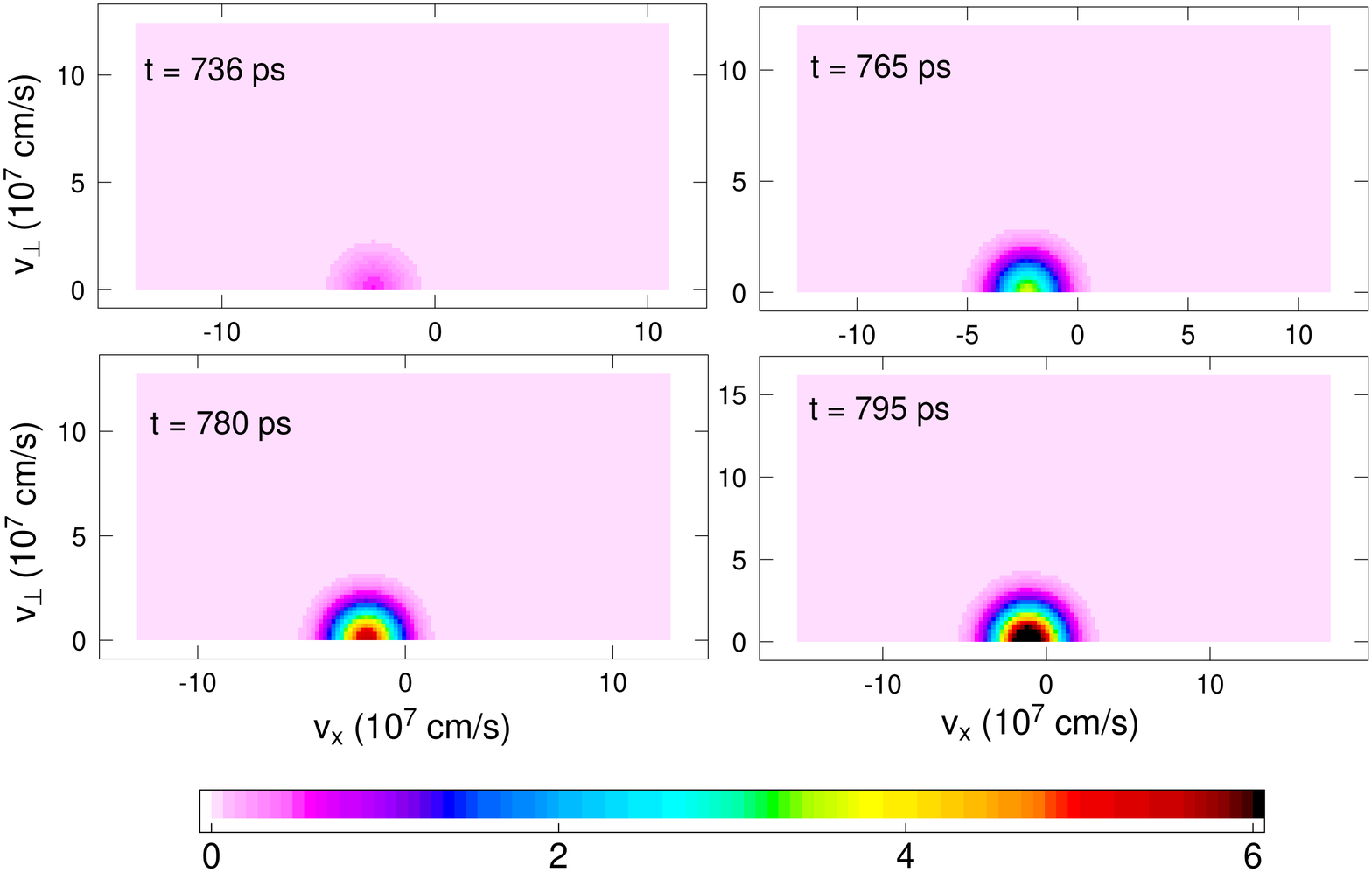}}
\caption{\label{fdcold} Thermal component of the $\alpha$ distribution function observed in a given cell of the imploding  hot spot at different times. This component is fed by the relaxation of the suprathermal component. The values of the distribution function are expressed in the units presented in table \ref{tabunit}. Times refer to beginning of the kinetic calculation.
}
\end{figure}

When the slowed down suprathermal $\alpha$-particles reach the thermal velocity region, a fraction of $\alpha$-particles is removed from the suprathermal component, to feed the thermal component according to Eq.\,\eqref{eq:eqFP_t}. The sequences represented in figures \ref{fdhot_1}- \ref{fdhot_2} illustrates this coupling from the suprathermal component point of view. The distribution function remains stable, while the particles are accumulating in the vicinity of the thermal region. Without the removal of the term \eqref{eq:third_term} on the right hand side of Eq.\,\eqref{eq:FP_supra_discr}, the suprathermal distribution function would have become unstable as $v\to V_0$. 
The evolution of the thermal component of the $\alpha$-particle distribution function  represented in figure\,\ref{fdcold}. It shows how the thermal component builds up.


\subsection{Ignition and burning wave propagation}

\begin{figure}[!h]
        \centering
        \begin{subfigure}[b]{0.5\textwidth}
                \rotatebox{-90}{\includegraphics[width=\textwidth]{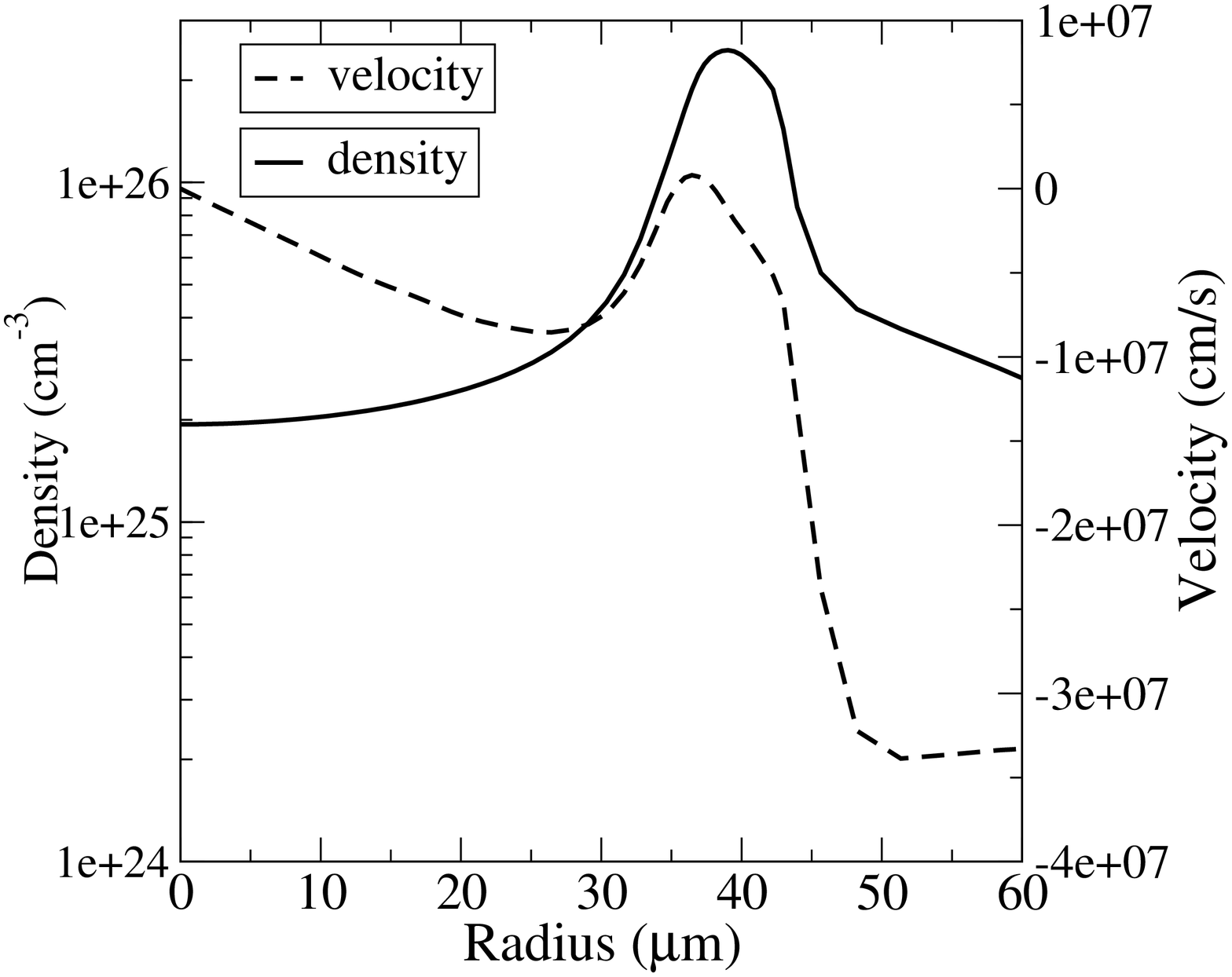}}
                \label{fig:gull}
        \end{subfigure}%
        ~ 
        \begin{subfigure}[b]{0.5\textwidth}
                \rotatebox{-90}{\includegraphics[width=\textwidth]{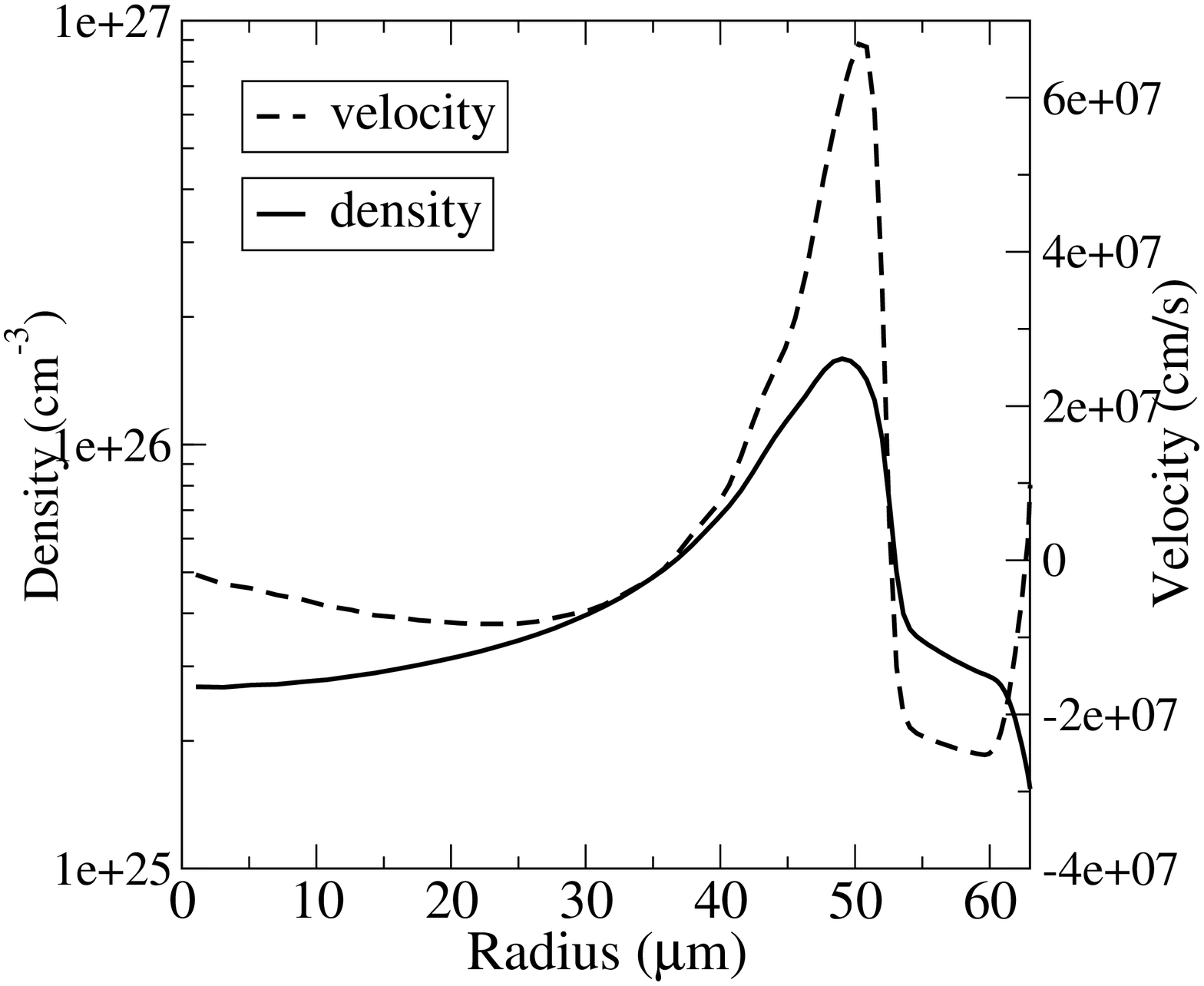}}
                \label{fig:tiger}
        \end{subfigure}
        ~ 
        \centering
        \begin{subfigure}[b]{0.5\textwidth}
                \rotatebox{-90}{\includegraphics[width=\textwidth]{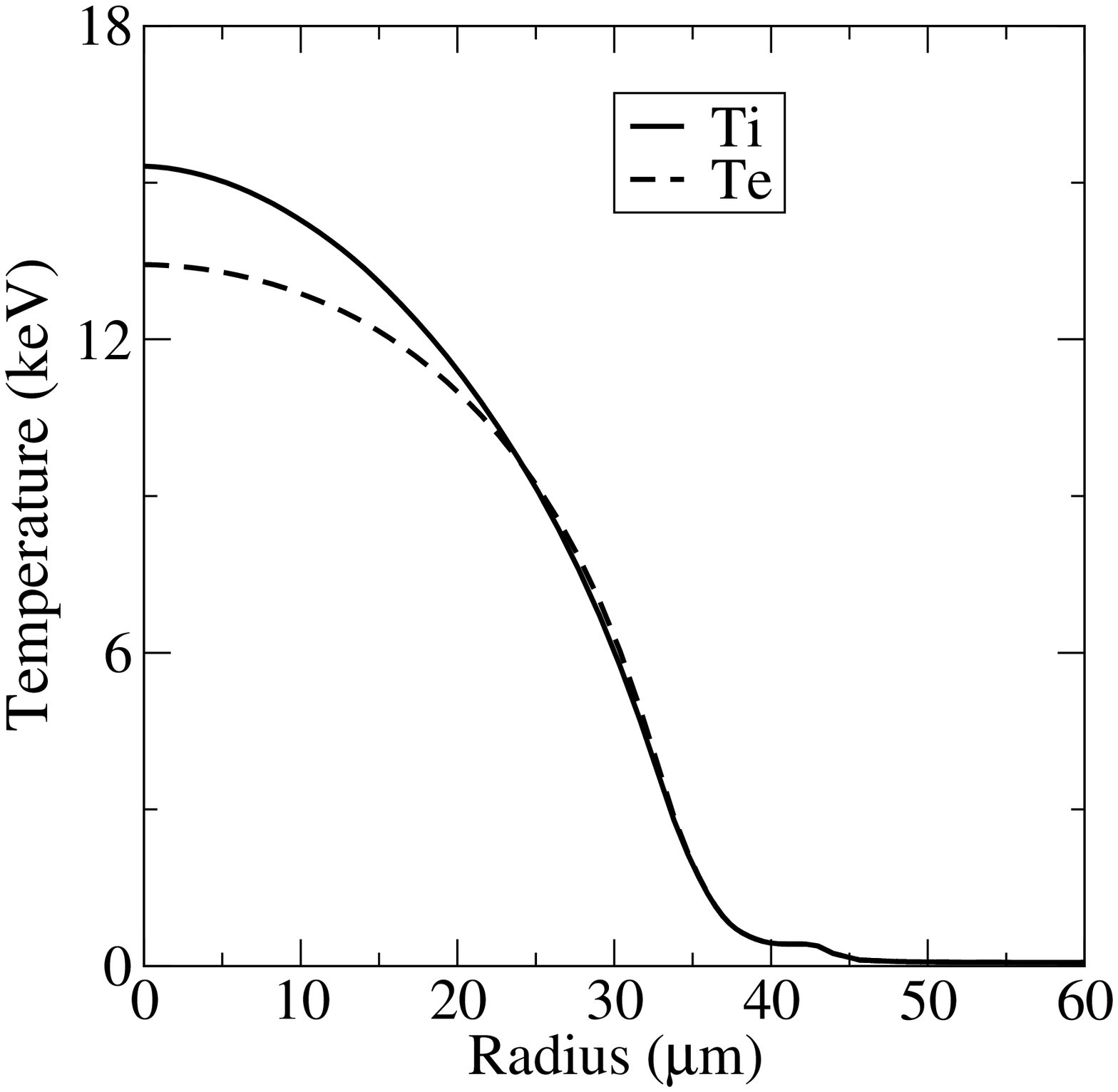}}
                \label{fig:mouse}
        \end{subfigure}%
        \begin{subfigure}[b]{0.5\textwidth}
                \rotatebox{-90}{\includegraphics[width=\textwidth]{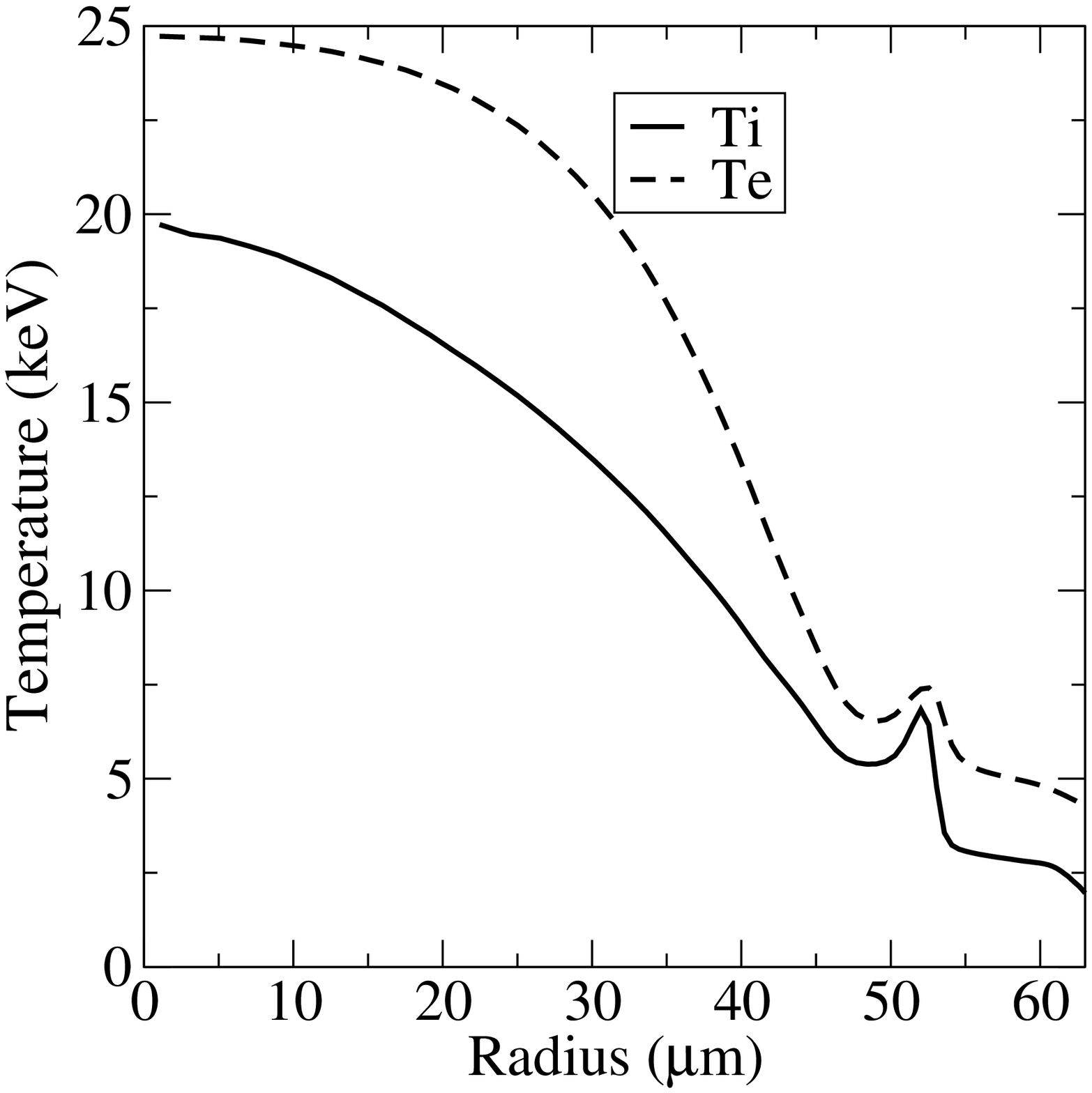}}
                \label{fig:mouse}
        \end{subfigure}
        \caption{\label{kinhydr850}
Profiles of the density ($n$), velocity ($u$) (top), of the electron ($T_e$) and total ($T_i$)
ion temperatures (bottom) in the fuel at $t=17.85$ ns in a DT ignition target
implosion; left : fluid simulation results, right : kinetic simulation results.}
\end{figure}

We finally give the density, velocity and temperature profiles calculated by \textsc{Fuse} and compare the results with the fluid code at the time $t=17.85$ ns (Fig.\ref{kinhydr850}) . After that time, corresponding to the arrival of the flame near the outermost cells, the kinetic simulation may not be relevant since the boundary condition (which comes from the hydrodynamic calculation) may not be consistent with the pressure calculated by the kinetic code. In the kinetic calculation, the heating of the hot spot appears to be faster than in the fluid code. This is consistent with the differences observed during the implosion phase, where the dense zone corresponding to the ablated cold fuel was imploding faster in the kinetic calculation. Besides, the kinetic ion temperature profile  displays a preheating wave ahead of the main temperature front. This is specially visible on the ion temperature profiles of (Fig.\ref{kinhydr850}). This structure is related to the Bragg peak of the D,T ions located in the dense cold fuel cold. Suprathermal $\alpha$-particles are created mainly in the central hot spot and deposit their energy and momentum near the inner interface of the cold fuel, where the thermal ion heating occurs. This interpretation will be examined more closely with future kinetic calculations of different target designs (that may be less efficient than the one considered here).
By applying the efficient algorithm (based on a 2-scale approach) exposed and validated in Sec.4 on real target configurations (that could not be solved analytically),  the code \textsc{Fuse} is able to simulate the fuel of real ICF targets at a kinetic level over a time corresponding to 1 ns after the start of the implosion. One thus models the ignition and the beginning of the burning wave propagation. Besides, by making use of a parallelization method of the collisional part of the code (which is possible since we can calculate the effect of collisions in each spatial cell independently from the others), it takes  less than 1 day of computation time, which is roughly twice as long as the usual simulations performed by \textsc{Fpion}  (corresponding to the implosion phase without $\alpha$-particles).


\section{Summary and  perspectives}\label{sec6}

We have  developed a numerical strategy to model fast $\alpha$-particles produced by fusion reactions at a ion kinetic level. A two-scale approach has been specially-tailored to represent the two-component nature of the $\alpha$ distribution function and simulate the thermalization process accurately.

Efficient algorithms have been designed to simulate the time evolution of the fast $\alpha$  component, driven by the transport in the inhomogeneous thermal plasma as well as the Coulomb collisional relaxation on electrons and ions.  The energy and momentum exchange between fast fusion products and the thermal plasma are thus calculated at the kinetic level. The methods have been tested in thermodynamic conditions corresponding to typical DT targets close to ignition. It has been shown that a locally split explicit scheme can be used to describe the fast $\alpha$ population evolution in non-prohibitive computational time. Besides, the algorithms presented here are easily parallelizable to take advantage of present-day multi-core architectures. 

The ion-kinetic code \textsc{Fuse}, built as an extension of the former code \textsc{FPion}, is thus able to model a full DT target implosion, including the ignition and burn processes, at a ion-kinetic level. Investigating in more details the role of kinetic effects of fusion products in the ignition and burn of DT targets is the purpose of ongoing work and will be published elsewhere \cite{future}. We may have in view to study implosions in the vicinity of the ignition threshold, where kinetic effects should be enhanced and may modify the energy gain. 

Finally, the algorithms developed here may be naturally extended to add the effect of Boltzmann-type large angle scattering, that would feed a suprathermal component for the D,T ions. Neutron momentum and energy deposition may be modeled in a similar way. 

\textbf{Acknowledgments.}
The authors are grateful to Professors Xavier Blanc, Josselin Garnier, R\'{e}mi Sentis and Gerald Samba for fruitful discussions on the subject.

\end{document}